\documentclass[final,5p,twocolumn]{elsarticle}

\usepackage{amsmath}
\usepackage{lineno,hyperref}

\hypersetup{
  colorlinks   = true, 
  urlcolor     = blue, 
  linkcolor    = blue, 
  citecolor   = red 
}

\begin{document}

\begin{frontmatter}

\title{{\bf Vertical Bifacial Solar Farms: Physics, Design, and Global Optimization}}

\author[purdue]{M. Ryyan Khan  \fnref{eq_contr} }
\author[kaust]{Amir Hanna  \fnref{eq_contr} }
\author[purdue]{Xingshu Sun  \fnref{eq_contr} }
\author[purdue]{Muhammad A. Alam\corref{email_Alam}\fnref{eq_contr}}\ead{alam@purdue.edu}

\address[kaust]{Electrical Engineering Department, King Abdullah University of Science and Technology, Saudi Arabia}

\address[purdue]{School of Electrical and Computer Engineering, Purdue University, West Lafayette, IN 47906, United States}

 \fntext[eq_contr]{The authors contributed equally.} 
\cortext[email_Alam]{Corresponding author}

\begin{abstract}
There have been sustained interest in bifacial solar cell technology since 1980s,  with prospects of 30-50\% increase in the output power from a stand-alone single panel.  Moreover,  a  vertical bifacial panel reduces dust accumulation and provides two output peaks during the day, with the second peak aligned to  the peak electricity demand. Recent commercialization and anticipated growth of bifacial panel market have  encouraged a closer scrutiny of  the integrated power-output and economic viability of bifacial solar farms, where mutual shading will erode some of the anticipated energy gain associated with an isolated, single panel. Towards that goal, in this paper we focus on geography-specific optimizations of {\it ground-mounted  vertical bifacial solar farms  for the entire world.} For local irradiance, we combine the  measured meteorological data  with the clear-sky model. In addition, we consider the detailed effects of direct, diffuse, and albedo light. We assume the panel is configured into sub-strings with bypass-diodes. Based on calculated light collection and panel output,  we analyze the optimum farm design  for maximum yearly output  at any given location in the world. Our results predict that,   regardless of the geographical location, a vertical bifacial  farm will yield 10-20\% more energy than a traditional monofacial farm for a practical row-spacing of 2m (1.2m high panels). With the prospect of additional 5-20\% energy gain from reduced soiling and tilt optimization, bifacial solar farm do offer a viable technology option for large-scale solar energy generation.  

\end{abstract}


\end{frontmatter}


\section{Introduction}

A conventional monofacial panel collects light only from the front side; the opaque back-sheet prevents collection of light scattered from ground (or surroundings) onto the back face of these panels. This extra energy from albedo can be partially recovered using a bifacial panel, where both faces of the panel and the cells are optically transparent. The concept of bifacial panels have been analyzed and experimentally demonstrated since 1980s \cite{luque_double-sided_1980,cuevas_50_1982}. Indeed, an isolated  bifacial panel has been shown to have up to 50\% extra output \cite{cuevas_50_1982} compared to a monofacial panel. Moreover, recent improvements in the design and fabrication of  bifacial cell technology suggest several additional advantages \cite{guerrero-lemus_bifacial_2016}. For example,  bifacial cells have a lower operating temperature (absence of infrared absorption at back metal) and better temperature coefficient  (e.g., HIT cells)---which would improve lifetime and integrated power output.   

Several studies in the literature have reported energy output of isolated, {\it standalone} bifacial panels both numerically \cite{luque_diffusing_1985, chieng_computer_1993, yusufoglu_analysis_2015, appelbaum_view_2016} and experimentally \cite{janssen_outdoor_2015, castillo-aguilella_multi-variable_2016,lave_performance_2016}. These studies include optimization of the tilt angle and elevation from ground for a single bifacial panel at various locations in the world. The recent work by Guo \textit{et al.} \cite{guo_vertically_2013} provides a global analysis of vertical bifacial panel. Given an albedo threshold, they have shown that an isolated  vertical panel will always produce more power compared to  an optimally tilted monofacial panel, irrespective of the geographic location. 

The energy gain of an isolated panel defines the upper limit of the performance potential of a solar cell technology.  Eventually, the panels will have to be installed in a farm, where one must account for the mutual shading of the panels. Clearly, the area-averaged power output will now be reduced. Under these circumstances, it is not clear if the advantages found for isolated panels can still sustain.  Recently,  Appelbaum \cite{appelbaum_bifacial_2016} has provided a partial answer by analyzing a solar farm at  Tel-Aviv (latitude $32^\circ$N).  His work focused on vertically vs. optimally tilted bifacial panel arrays. The  optimally tilted farm yields 32\% more energy than the vertical farm (in latitude $32^\circ$N)---however, it is not clear how the outputs compare to the monofacial panel array.   It is also difficult to know if the conclusions apply to other regions of the world. An analysis that broadens the previous work to all the locations of the world (a global optimization) will be helpful. This analysis is particularly important because ITRPV roadmap projects that the bifacial market share will increase from 5\% in 2016 to 30\% in 2026 \cite{noauthor_international_2016}. Many PV manufacturers (e.g., Panasonic, Prism Solar, LG, SolarWorld, Centrotherm, etc.) are now producing bifacial panels. A few recent solar farms (e.g., Asahikawa Hokuto Solar Power Plant in Japan, and La Silla PV plant in Chile) are utilizing bifacial panels. Given this rapid progress, it is important to clearly understand the complex physics, design, and optimization  of bifacial solar farms.

Among various farm configurations, vertically aligned bifacial panels have been of particular interest because of reduced soiling (dirt or snow) which increases overall energy output. In addition, the higher output in the afternoon due to the `double-humped' daily output profile \cite{guo_vertically_2013} coincide with the peak electricity demand. Since optimally tilted bifacial panels will always produce slightly more energy compared to the vertical farms, the analysis of vertically aligned panels may be viewed as a lower limit of energy produced by an optimized bifacial farm.

In this paper, we offer detailed model, physics, and a worldwide perspective regarding ground-mounted vertical bifacial solar farms.

We combine the global meteorological data from NASA with the clear-sky model from Sandia to estimate hourly insolation. This new algorithm bypasses the loading of extremely large hourly database, and allows efficient computation towards global analysis of new technologies while maintaining realistic and daily averaged meteorological information.

Next, we model the direct and diffused light collection \cite{passias_shading_1984,bany_effect_1987,doubleday_recovery_2016}, as well as the non-trivial physics of albedo light collection \cite{fathi_view_2016,appelbaum_view_2016} while accounting for relevant shadings on the panels and the ground. Our generalized formulation models the  non-uniform illumination along the panel height. Only a fraction of the light incident on the panels will produce electricity \cite{sanchez_reinoso_simulation_2013} because of the spatially non-uniform illumination and the nature of the electrical connection for the panels. The second aspect is often not accounted for in literature. We use the spatially non-uniform light collection data along with the appropriate circuit model of the panels to accurately find the hourly \textit{energy-output} from the panels and the farm.  

Mutual shading between adjacent panels penalizes energy-output, thereby restricting panels from being closely packed in the farm. We explain how this results in an optimum period between the panels. At high latitudes, the sun-path is more tilted, resulting in larger optimum panel-period. In addition, at the same latitude, locations with more diffuse insolation tend to have a larger panel-period.

Finally, we present a global perspective on the annual yield of vertical bifacial solar farms. The key conclusion of the paper is this: With inter-row separation of 2m (typically required for maintenance) for 1.2m wide panels, a ground-mounted vertical bifacial farm  outperforms a traditional monofacial farm by 10-20\%, regardless of the geographical location. The gain may persist even for smaller inter-row separation, once the energy loss due to soiling \cite{adinoyi_effect_2013,bouaouadja_effects_2000,lu_numerical_2016, menoufi_dust_2017} is accounted for. The performance gain requires a denser packing of vertical bifacial panels, the implication of which must be accounted for in the levelized cost of electricity (LCOE) calculation \cite{janko_implications_2016,lai_levelized_2017}. 

In sections \ref{sec:irradModel} and \ref{sec:physicsModel}, we present the details of the irradiance model, and the physical model to calculate the light collection and power generation of the panels and the farm. In sections \ref{sec:physicsResult}-\ref{sec:kTResult}, we discuss the physics and design-optimization of the farm. Finally, in section \ref{sec:globalResult}, we present the global perspective and prospects of the optimally designed vertical bifacial solar farm. Our conclusions are summarized in section \ref{sec:conclusion}.

\section{Method}

\subsection{Irradiance model}
\label{sec:irradModel}

\noindent\textbf{2.1.1 Simulation of hourly GHI.} Temporal solar irradiance data  consist of  the position of the sun and its intensity. This  information is crucial to simulate and optimize the energy yield of solar farms.

To simulate such data, we first start by calculating the position of the sun (solar Zenith $\theta_Z$ and Azimuth  $\gamma_S$ angles) at arbitrary time and geographic locations by using the NREL's solar position algorithm \cite{reda_solar_2004} implemented in Sandia model library \cite{sandia_national_labs_pv_2016}. Here, $\theta_Z$ is the refraction-corrected Zenith angle, which depends on altitude and ambient temperature.
Second, we input the sun position data into the Haurwitz clear sky model to generate the Global Horizontal Irradiance (GHI or $I_{GHI}$) \cite{haurwitz_insolation_1945,haurwitz_insolation_1946} on a minute-to-minute basis. 
Note that the clear sky model often overestimates insolation, especially when the atmosphere is cloudy or overcast. Hence, in the third and final step, we integrate the simulated GHI over time, which is then scaled to match the satellite-derived monthly average GHI data (for 22 years) from the NASA Surface meteorology and Solar Energy database \cite{power_surface_2017}, whereby local variation of GHI caused by cloudiness and altitude is incorporated into the calculation. Therefore, our modeling framework fully incorporates the impacts of geographic and climatic factors into modeling the solar irradiance.\\


\begin{figure}[t]
\vspace{-0pt}
\centering
\includegraphics[width=0.57\textwidth]{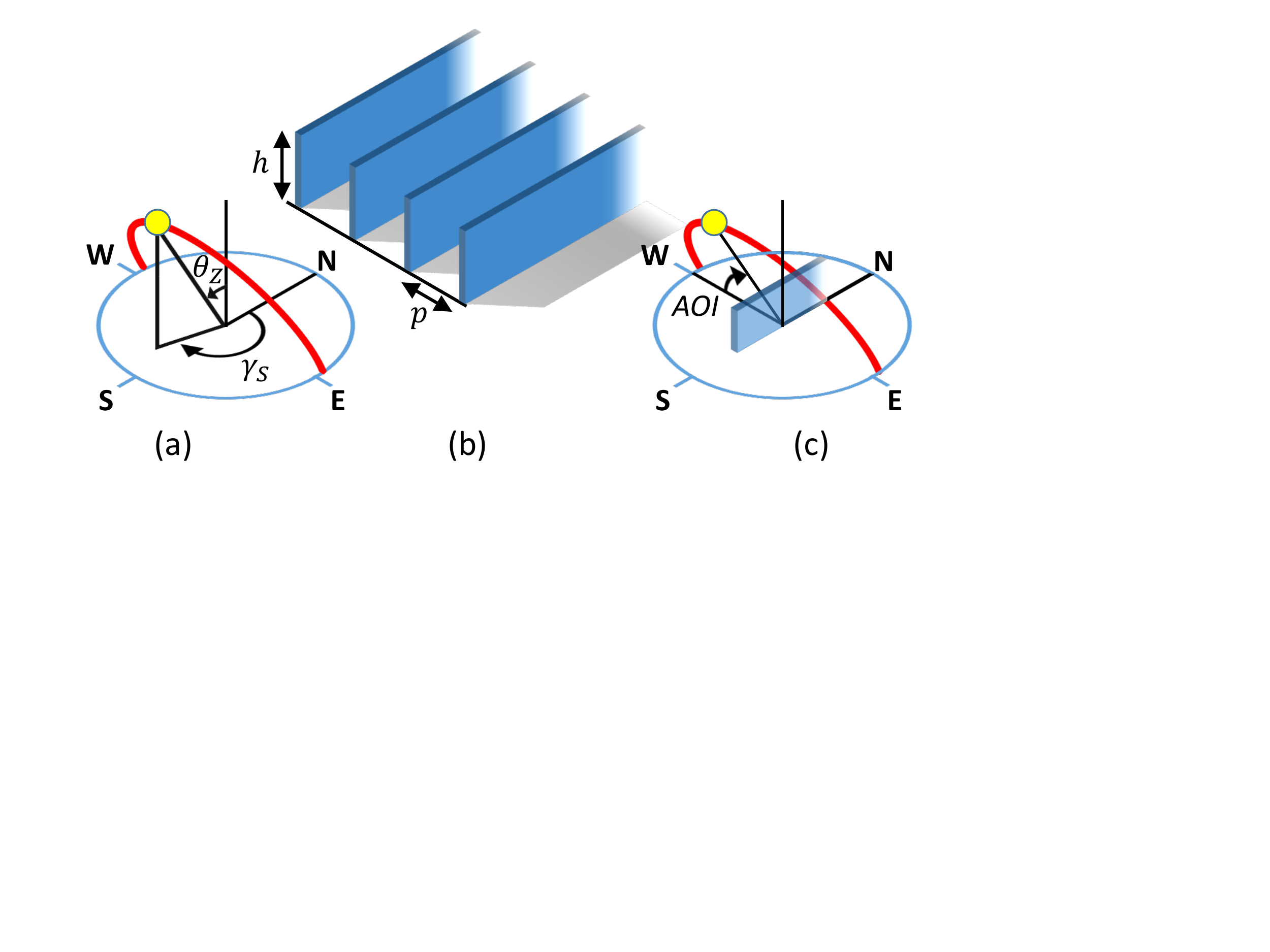}
\vspace{-130pt}
\caption{ (a) The zenith and azimuth angles ($\theta_Z$, $\gamma_S$) of the sun is shown at a specific time. An example of the sun-path is shown by the red line. (b) The vertical bifacial solar farm is depicted with relevant definitions. (c) This shows the angle of incidence (AOI) on the panel at the specific solar position.  }
\label{fig:definitions}
\end{figure}


\noindent\textbf{2.1.2 Decomposition of  GHI into DHI and DNI.} Calculating the irradiance on a tilted surface requires decomposing GHI into  two components: Direct Normal Irradiance (DNI or $I_b$) and Diffuse Horizontal Irradiance (DHI or $I_{diff}$).  The relationship between the two components  can be written as
\begin{equation}
	I_{GHI}=I_b \cos\theta_Z + I_{diff}. \label{eq:def_GHI}
\end{equation}
Based on (\ref{eq:def_GHI}), however, it is impossible to separate $I_b$ and $I_{diff}$ from $I_{GHI}$. Therefore, we estimate the diffuse fraction of $I_{GHI}$ using the Orgill and Hollands model which empirically calculate the diffuse fraction using the clearness index of the sky ($k_T$) \cite{orgill_correlation_1977}. The clearness index is defined as the ratio between $I_{GHI}$ and extraterrestrial irradiance ($I_0$) on a horizontal surface, i.e.,
\begin{equation}
	k_T=\frac{I_{GHI}}{I_0 \times \cos\theta_Z}.   \label{eq:def_kT}
\end{equation}  			                         
For a specific time and location, $I_{GHI}$ is already known while the extraterrestrial irradiance can be evaluated analytically \cite{duffie_solar_2013}; therefore, we can obtain the clearness index $k_T$ on a minute-to-minute basis using (\ref{eq:def_kT}). Knowing $I_{GHI}$ and $k_T$, we use the Orgill and Hollands model to determine $I_{diff}$, which allows us to deduce $I_b$ from (\ref{eq:def_GHI}). For demonstration, an example calculation of irradiance  at Washington DC on September 22 is shown in Fig. \ref{fig:daily_power}.  

There are several empirical models for decomposing GHI found in literature \cite{maxwell_quasi-physical_1987,erbs_estimation_1982,reindl_diffuse_1990}. Generally, good agreement have been found among these models \cite{wong_solar_2001}. Also, we assume isotropic sky model \cite{loutzenhiser_empirical_2007} for diffuse irradiance $I_{diff}$. The Perez model \cite{perez_dynamic_1992} provide a more elaborate and somewhat more complex representation of the diffuse light. However, we expect that our numerical results and key conclusion will not be overly sensitive to the model chosen, and the general conclusions will hold irrespective of the assumed models.\\

\noindent\textbf{2.1.3 Angle of Incidence (AOI) calculation.} 
To evaluate the contribution of the beam component of sunlight (i.e., $I_b$), we need to calculate the angle of incidence (AOI) between $I_b$ and the front/back surface of vertical bifacial solar panels. It turns out that AOI of an east-west facing vertical bifacial solar panel can be simply expressed as
\begin{align}
	\theta^{(F)} &= AOI_{front} = \cos^{-1} \left[ \sin\theta_Z \times \cos(\gamma_S -\pi/2) \right], \\
    \theta^{(B)} &= AOI_{back} = \cos^{-1} \left[ \sin\theta_Z \times \cos(\gamma_S +\pi/2) \right]. 
\end{align}
for the front and back surfaces, respectively. In the next section, knowing the angular and irradiance data of sunlight, we will show how to evaluate the optical absorption and power generation of vertically-mounted bifacial solar farms.

\subsection{An array collects direct, diffuse, and albedo light}
\label{sec:physicsModel}

The solar farm consists of  vertical bifacial panels of height $h$, separated by a period of $p$, as shown in Fig. \ref{fig:definitions}(b). Each of the panels face E-W and run infinitely along N-S direction. The front face (East facing) sees the sun from sunrise until noon. The back face (West facing) of the panel sees the sun from noon until sunset. In the following, we will first explain light collection by individual panels and we will integrate the contributions from the array to calculate total power output from the farm. \\

\noindent \textbf{2.2.1 Panel properties: uniform illumination.}
Let us assume the panels have monofacial-efficiency of $\eta$ for uniform, normal illumination  onto the panel. For an angle of incidence (AOI) $\theta$, we can approximate the efficiency as $\eta(\theta) \equiv [1-R(\theta)]\times \eta$. The angle dependent reflectivity of the panel can be empirically written as \cite{martin_annual_2005}
\begin{equation}
	R(\theta) = 1- \frac{1-\exp(-\cos\theta /a_r)}{1-\exp(-1 /a_r)}.
\end{equation}
Here, $a_r$ is the angular loss coefficient. In the following calculations, we assume $a_r=0.16$, typical for commercial Si solar panels \cite{martin_annual_2005}.

The efficiency $\eta_{\textit{diff}}$ of the panel under diffuse sunlight (isotropic illumination) will be lower than that under normal (direct) illumination. We assume $\eta=18.9\%$ and $\eta_{\textit{diff}}=15.67\%$ under normal and diffuse illumination on a single face of the panel (estimated using the simulator `Tracey' \cite{keith_r._mcintosh_tracey_2013,mcintosh_optical_2009,mcintosh_optical_2010}: see Supplementary Information (SI)). Oblique angles in the diffuse light have higher reflection loss than normal incidence---that is why $\eta_{\textit{diff}} < \eta$.

Experimentally, the cell efficiency observed by illuminating front or the back faces differ by 1-2\% \cite{guerrero-lemus_bifacial_2016}. For simplicity, we assume these values to be the same. Our formulation is general, and can be applied for the bifacial efficiency asymmetry by using separate values of $\eta^{(F)}$ and $\eta^{(B)}$ for front and back face efficiencies, respectively.  

At any specific time of the day, the two faces of a bifacial panel are illuminated asymmetrically. Therefore, we  calculate the power collection from the front and back faces separately. Let us assume that at any given time of the day,  AOI for the front and back panel faces are  $\theta^{(F)}$ and $\theta^{(B)}$ ,  respectively. We will focus on the power collection from the front face, and the calculations for the back face will follow a similar approach. We will denote power per unit area of the \textit{panel-surface} and per unit area of the \textit{farm-land} by $\widehat{I}_{PV(*)}$ and ${I}_{PV(*)}$, respectively. \\

\noindent\textbf{2.2.2 Panel properties: non-uniform illumination.}
During mornings and afternoons, mutual shadowing makes the illumination over the  panel  spatially non-uniform, with the lower part of the panel receiving less light than the top.   For a panel constructed from a set of series connected cells, bypass diodes are placed across different sub-sections of the series-string to avoid reverse breakdown of the shaded cells. We assume $N_{\textit{bypass}}=3$ bypass diodes sub-divide each panel into  $N_{\textit{bypass}}=3$ strings. The effect of the shading on lowering the panel output is taken into account based on the analytical approach developed by C. Deline \textit{et al.} \cite{deline_simplified_2013}. In this calculation, we assume that the total current is always limited by the bottom string; the validity of the assumption is discussed in the SI-document.\\


\begin{figure}[t]
\vspace{-0pt}
\centering
\includegraphics[width=0.65\textwidth]{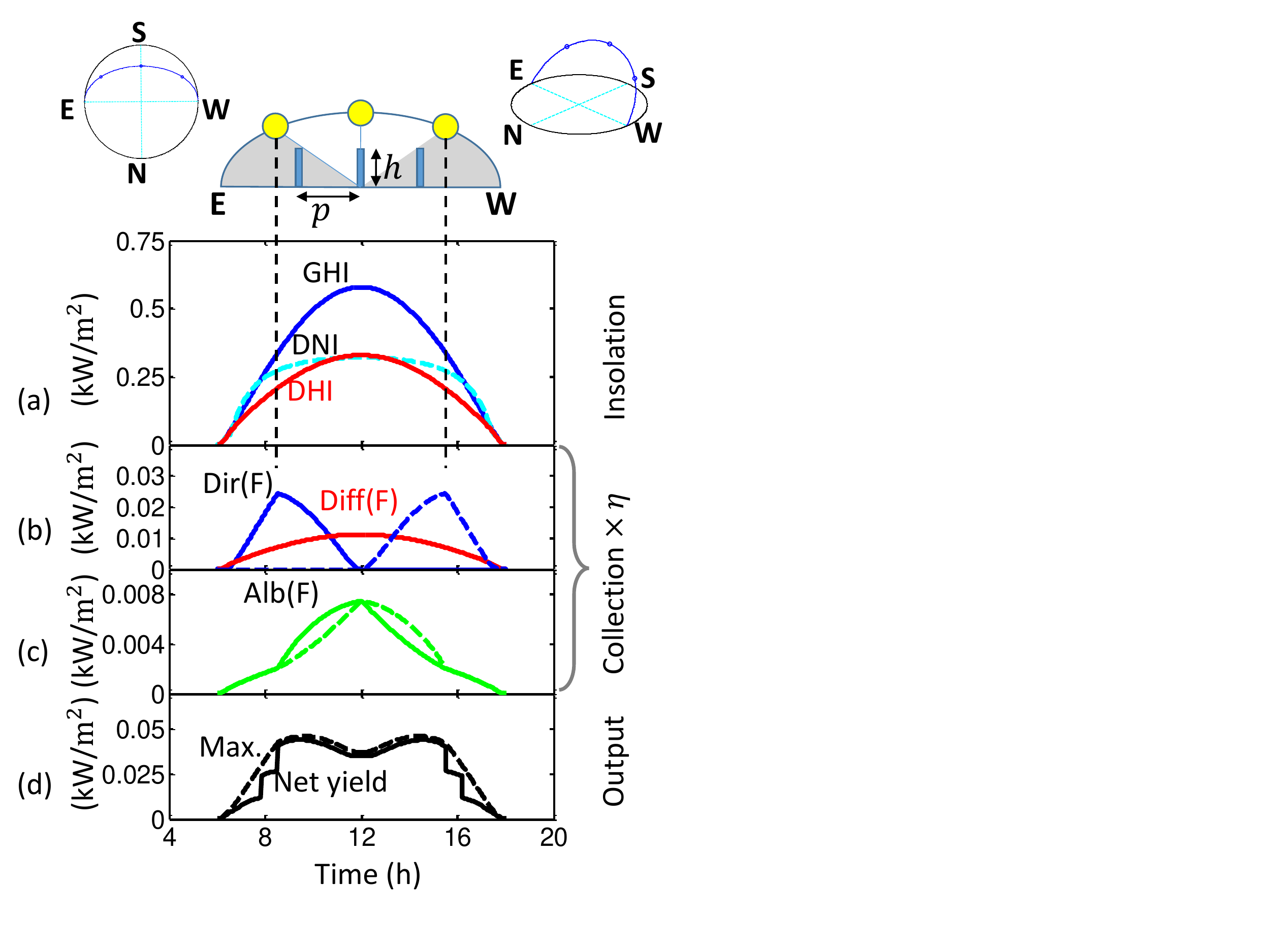}
\vspace{-40pt}
\caption{ (a) Hourly variation of insolation components for Washington DC on September 22. (b, c, d) The power generation components of the farm is shown. Here, we have assumed $h=1.2$m and $p=2$m.  }
\label{fig:daily_power}
\end{figure}


\noindent\textbf{2.2.3 Direct insolation collection.}
At any given time of the day, direct illumination component normal to the panel is $I_b \cos\theta^{(F)}$ for the unshaded part of the panel (i.e., $z>h_s$), see Fig. \ref{fig:result_opt}(c). Here, $z$ is any position along the height on the panel, and $h_s$ is the shadow on the panel  \cite{passias_shading_1984,bany_effect_1987} the corresponding time of the day. Considering the reflection loss $R(\theta^{(F)})$ and the panel efficiency $\eta^{(F)}$, we  find the power generated per surface area at height $z$ of the front face of the panel as follows,
\begin{equation}
	\widehat{I}_\textit{PV(dir)}^{(F)}(z) =  
\begin{cases}
 [1-R(\theta^{(F)})]\eta^{(F)} I_b \cos\theta^{(F)},  & z>h_s  \\
0,    & z \le h_s .\\
\end{cases} \label{eq:Ipv_dir_z}
\end{equation}
Here, $\widehat{I}_\textit{PV(dir)}^{(F)}(z)$ is the power generation component only from the direct/beam sunlight. The corresponding integrated `maximum' power (\textit{per unit farm area}) from the direct sunlight is given by:
\begin{align}
	I_\textit{PV(dir),0}^{(F)} &= \frac{1}{p} \times \int_0^h {\widehat{I}_\textit{PV(dir)}^{(F)}(z)\, dz} \nonumber \\
    & = \frac{(h-h_s)}{p}[1-R(\theta^{(F)})]\eta^{(F)} I_b \cos\theta^{(F)} . \label{eq:Ipv_dir_0}
\end{align}
We quote $I_\textit{PV(dir),0}^{(F)}$ as the `maximum' output from direct light as this does not assume any loss due to non-uniform generation in the series connected string of cells. This maximum may be reached, for example, in a thin-film like panel configuration. The solid line in Fig. \ref{fig:daily_power}(b) shows $I_\textit{PV(dir),0}^{(F)}$ as the day progresses. After the solar-noon, the front face will not directly see the sun, therefore  $I_\textit{PV(dir),0}^{(F)}=0$ for the later part of the day. Similarly, the back face shows a mirrored characteristic for $I_\textit{PV(dir),0}^{(B)}$ as shown by the dashed line. These two components together contribute to the characteristic double-humped hourly output profile of a vertical bifacial panel. \\


\begin{figure}[t]
\vspace{-0pt}
\centering
\includegraphics[width=0.60\textwidth]{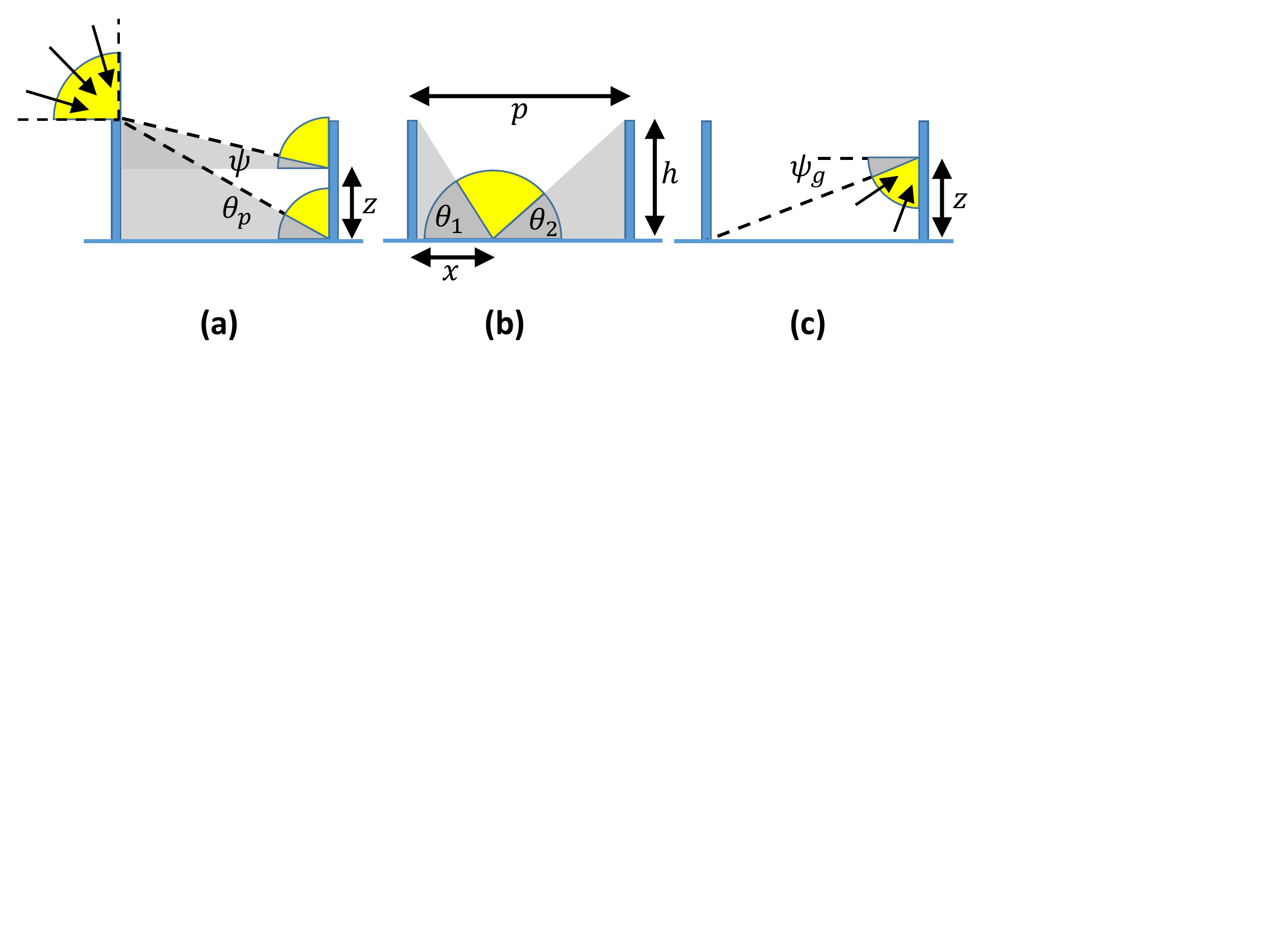}
\vspace{-160pt}
\caption{(a) Partial masking of DHI on a face of the panel. (b) Partial masking of DHI on the ground. The fractional DHI reaching the ground is a source for albedo light. (c) Collection of the above mentioned albedo light. }
\label{fig:Ch30_solarFarm-4}
\end{figure}


\noindent\textbf{2.2.4 Diffuse insolation collection.}
Ideally, when the panels are far apart, half of the diffuse rays angle towards the front-face of the panel. These rays cover zenith angle range of $[-\pi/2,0]$. However, a fraction of these angles is obstructed/shaded when the panels are arranged in an array---depicted by the shaded quarter circles in Fig. \ref{fig:Ch30_solarFarm-4}(a). In this illustration, we  see that the top portion of the vertical panel receives more diffuse light than the bottom. Calculation of diffuse insolation collection using the well-known average diffuse masking angle \cite{passias_shading_1984} will yield in an overestimated value, especially for highly tilted panels. An appropriate view factor \cite{appelbaum_bifacial_2016} properly estimates the diffuse light collection. 

The incident diffuse light intensity at height $z$ of the panel (see Fig. \ref{fig:Ch30_solarFarm-4}(a)) is $I_{diff}\times F_{\textit{dz-sky}}$. The diffuse light is masked at angle $\psi(z)$ resulting in the view-factor (towards the unobstructed sky) of  $F_{\textit{dz-sky}}=(1-\sin\psi(z))/2$ at $z$ \cite{modest_radiative_2013}. The corresponding power generation per panel area (front face) from the diffuse insolation is as follows,
\begin{align}
	\widehat{I}_\textit{PV(diff)}^\textit{(F)}(z) 
    &= \eta_{\textit{diff}}^{(F)} 
    	\left[ I_{\textit{diff}} \times F_{\textit{dz-sky}} \right] \nonumber \\ 
    &= \eta_{\textit{diff}}^{(F)} 
    	\left[ I_{\textit{diff}} \times \frac{1}{2}(1-\sin\psi(z)) \right].\label{eq:Ipv_diff_z}
\end{align}
And, the corresponding integrated, `maximum' power generation \textit{per unit farm area} from the diffuse light is:
\begin{align}
	I_\textit{PV(diff),0}^\textit{(F)} 
    &= \frac{1}{p}\times \int_0^h{\widehat{I}_\textit{PV(diff)}^\textit{(F)}(z) \, dz}  \nonumber \\
    &=\frac{h}{p} \, \eta_{\textit{diff}}^{(F)} 
    	\left[ I_{\textit{diff}} \times \frac{1}{2}(1-\tan(\theta_p/2)) \right]. \label{eq:Ipv_diff_0}
\end{align}
Here, $\theta_p=\tan^{-1}(h/p)$. The hourly variation of $I_\textit{PV(diff),0}^\textit{(F)}$ is shown by the red solid line in Fig. \ref{fig:daily_power}(b). As expected, this component of power generation peaks at noon when the DHI ($I_{diff}$) also peaks. In the above calculations, we can find $\widehat{I}_\textit{PV(diff)}^\textit{(B)}(z)$ and $I_\textit{PV(diff),0}^\textit{(B)} $ for the back face by replacing $\eta_{\textit{diff}}^{(F)}$ with $\eta_{\textit{diff}}^{(B)}$.\\

\noindent \textbf{2.2.5 Albedo light collection.}
Let us first describe the effect of diffuse insolation on albedo. As explained in the preceding discussion, there is a fractional-shadowing (or masking) of the diffuse light reaching the panel. A similar scenario is true for diffuse light reaching the ground. And, depending on the position between the panels, the amount of diffuse sunlight reaching the ground is different. 

Consider a position $x$ between adjacent panels, as in Fig. \ref{fig:Ch30_solarFarm-4}(c). The masking angles from the two panels are:
\begin{equation}
	\theta_1(x) = \tan^{-1}\frac{h}{x} \quad \mbox{  and, } \theta_2(x)=\tan^{-1}\frac{h}{p-x} .
\end{equation}
The average masking angles can be written as,
\begin{align}
	\bar{\theta_1} &= \frac{1}{p} \int_0^p {\theta_1(x) \,dx} 
    	= \theta_p + \frac{\ln(\csc\theta_p)}{\cot\theta_p}.
\end{align}
Due to symmetry: $\bar{\theta_1}=\bar{\theta_2}$. Here, $\theta_p=\tan^{-1}(h/p)$. The average diffuse insolation reaching the ground is,
\begin{equation}
	I_{\textit{Gnd:diff}} = I_{\textit{diff}} \times \frac{1}{2} (\cos\bar{\theta_1}+ \cos\bar{\theta_2}) 
    =  I_{\textit{diff}} \times \cos\bar{\theta_1} .
\end{equation}
Note that $\bar{\theta_1}$ is constant throughout the day, and $I_{\textit{Gnd:diff}} $ is proportional to $I_{\textit{diff}} $.  Diffuse masking on the ground has not been considered in prior literature,  although the contribution is particularly important for typical $p/h$. For example, with $p/h\sim 1$, the DHI $I_{\textit{diff}} $ may be masked more than 50\% (i.e., $\cos\bar{\theta_1}<0.5$).  Now, $I_{\textit{Gnd:diff}}\times R_A $  can be the diffused light source for the front (or back) face of the panel. The albedo light collection originating from diffuse insolation is masked at angle $\psi_g(z)$ at height $z$ on the panel, see Fig.~\ref{fig:Ch30_solarFarm-4}(c). Therefore, the corresponding power generation per panel front surface at $z$ is:
\begin{align}
	\widehat{I}_\textit{PV(Alb:diff)}^{(F)}(z)
            &= \eta_{\textit{diff}}^{(F)} \, I_{\textit{Gnd:diff}} R_A \times F_\textit{dz-gnd} \nonumber \\
            &= \eta_{\textit{diff}}^{(F)} \, I_{\textit{Gnd:diff}} R_A \times \frac{1}{2} (1-\sin\psi_g(z)). \label{eq:Ipv_albDiff_z}
\end{align}
 The corresponding integrated, `maximum' power generation \textit{per unit area} is given by:
\begin{align}
	I_\textit{PV(Alb:diff),0}^{(F)}& = \frac{1}{p} \int_0^h{\widehat{I}_\textit{PV(Alb:diff)}^{(F)}(z)\, dz} \nonumber  \\
            &= \frac{h}{p}  \, \eta_{\textit{diff}}^{(F)} \, I_{\textit{Gnd:diff}} R_A \times \frac{1}{2} \left(1-\tan(\theta_p/2) \right) . \label{eq:Ipv_albDiff_0}
\end{align}


\begin{figure}[t]
\vspace{-0pt}
\centering
\includegraphics[width=0.55\textwidth]{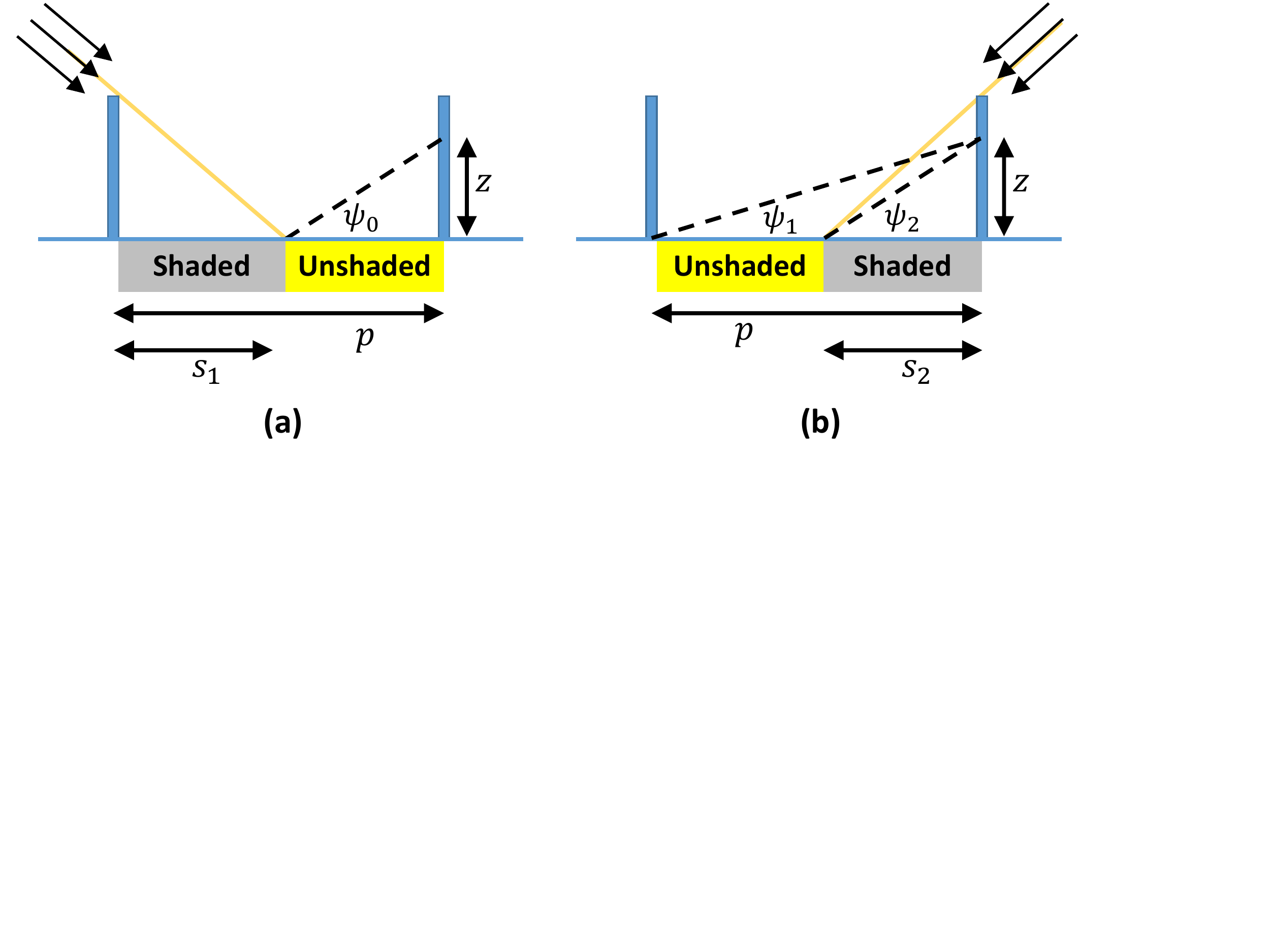}
\vspace{-130pt}
\caption{ Partial shading of the ground due to DNI during (a) morning, and (b) afternoon.   }
\label{fig:Ch30_solarFarm-5}
\end{figure}


Next, we can consider albedo from the direct insolation \cite{fathi_view_2016}. In the morning, the shading on the ground will be configured as shown in Fig. \ref{fig:Ch30_solarFarm-5}(a). The shade length $s_1$ is equal to the period $p$ (i.e., ground fully shaded for the beam component) when the array is turned on. The shade goes away ($s_1=0$) at noon. At any time of the day, the unshaded length $(p-s_1)$ subtends angles $[\psi_0,\pi/2]$ at the point $z$ on the front panel face (see Fig. \ref{fig:Ch30_solarFarm-5}(a)). In the afternoon, shading $s_2$ is adjacent to the front face, Fig. \ref{fig:Ch30_solarFarm-5}(b). In this case, the unshaded region subtends angles $[\psi_1,\psi_2]$ at the point $z$ on the front panel face. We can write:
\begin{align}
	\psi_0(z) &= \cot^{-1} \left(\frac{p-s_1}{z} \right), \nonumber \\ 
    \psi_1(z) &=\cot^{-1} \left(\frac{p}{z} \right) \mbox{ and } 
    \, 	\psi_2(z) = \cot^{-1} \left(\frac{s_2}{z} \right) .
\end{align}
The shadow length $s_1$ ( or $s_2$ ) is calculated for each time-step of the day \cite{passias_shading_1984,bany_effect_1987}. Finally, the power generated per area of panel front-face at $z$  from albedo originating from the direct sunlight is calculated as follows:
\begin{align}
	\widehat{I}_\textit{PV(Alb:dir)}^{(F)} (z)
        &=  \eta_{\textit{diff}}^{(F)} \, I_{\textit{dir}} R_A \times F_\textit{dz-Ugnd}, \label{eq:Ipv_albDir_z}
\end{align}
where, the view factor from the position $z$ on panel-face to the unshaded part of the ground is given by,
\begin{equation}
	F_\textit{dz-Ugnd} =  
\begin{cases}
\frac{1}{2} \left(1-\sin\psi_0(z) \right) , &  
				\mbox{ (before noon)} \\
\frac{1}{2} \left(\sin\psi_2(z)-\sin\psi_1(z) \right) , & \mbox{ (afternoon)} .\\
\end{cases}
\end{equation}
And, the corresponding integrated, `maximum' power generated \textit{per farm area} is:
\begin{align}
	I_\textit{PV(Alb:dir),0}^{(F)} &= \frac{1}{p} \int_0^h \widehat{I}_\textit{PV(Alb:dir)}^{(F)} (z)\, dz \nonumber \\
        &= \frac{h}{p} \, \eta_{\textit{diff}}^{(F)} \, I_{\textit{dir}} R_A \times F_\textit{PV-Ugnd}, \label{eq:Ipv_albDir_0}
\end{align}
where, the view factor from the full panel-face to the unshaded part of the ground is given by,
\begin{equation}
	F_\textit{PV-Ugnd} =  
\begin{cases}
\frac{1}{2} \left(1-\tan{\frac{\psi_0(h)}{2}} \right) , &  
				\mbox{ (before noon)} \\
\frac{1}{2} \left(\tan{\frac{\psi_2(h)}{2}} - \tan{\frac{\theta_p}{2}} \right) , & \mbox{ (afternoon)} .\\
\end{cases}
\end{equation}

The net `maximum' albedo light contribution from the front-face ($I_\textit{PV(Alb),0}^{(F)}=I_\textit{PV(Alb:dir),0}^{(F)}+I_\textit{PV(Alb:diff),0}^{(F)}$) is shown by the solid line in Fig. \ref{fig:daily_power}(c). For the back-face, $I_\textit{PV(Alb:dir),0}^{(B)} $ just the flipped version around noon.

Finally, combining Eqs. (\ref{eq:Ipv_dir_0}), (\ref{eq:Ipv_diff_0}), (\ref{eq:Ipv_albDiff_0}), (\ref{eq:Ipv_albDir_0}), the `maximum' net power generated per farm area is
\begin{align}
	I_\textit{PV,0}^{\textit{(bifacial)}} &=   
    \left[I_\textit{PV(dir),0}^{\textit{(F)}}+ I_\textit{PV(dir),0}^{\textit{(B)}} \right]+ 
\left[I_\textit{PV(diff)}^{\textit{(F)}}+ I_\textit{PV(diff),0}^{\textit{(B)}} \right]
\nonumber \\ & \qquad + \left[ I_\textit{PV(Alb),0}^{\textit{(F)}}+  I_\textit{PV(Alb),0}^{\textit{(B)}} \right] \\
    &=   I_\textit{PV(dir),0}^{\textit{(bifacial)}}+ I_\textit{PV(diff),0}^{\textit{(bifacial)}}+ I_\textit{PV(Alb),0}^{\textit{(bifacial)}} 
\end{align}

The black dashed line in Fig. \ref{fig:daily_power}(d) shows $I_\textit{PV,0}^{\textit{(bifacial)}}$ as the day progresses. Due to partial shading and non-uniform illumination, however, it is not possible to extract this power from the panel configured with the string and bypass-diode connection. A detailed calculation results in the final power generation $I_\textit{PV}^{\textit{(bifacial)}}$ per farm area, see black solid line in Fig. \ref{fig:daily_power}(d). The abrupt jumps correspond to times when the bypass diodes turns on or off certain sub-strings on the panel. Notice that $I_\textit{PV}^{\textit{(bifacial)}} < I_\textit{PV,0}^{\textit{(bifacial)}}$ throughout the day. The residual double-humped feature originates from the direct insolation component;  otherwise, the power-generation profile is flattened by diffuse and albedo light.

\begin{figure}[t]
\vspace{-0pt}
\centering
\includegraphics[width=0.5\textwidth]{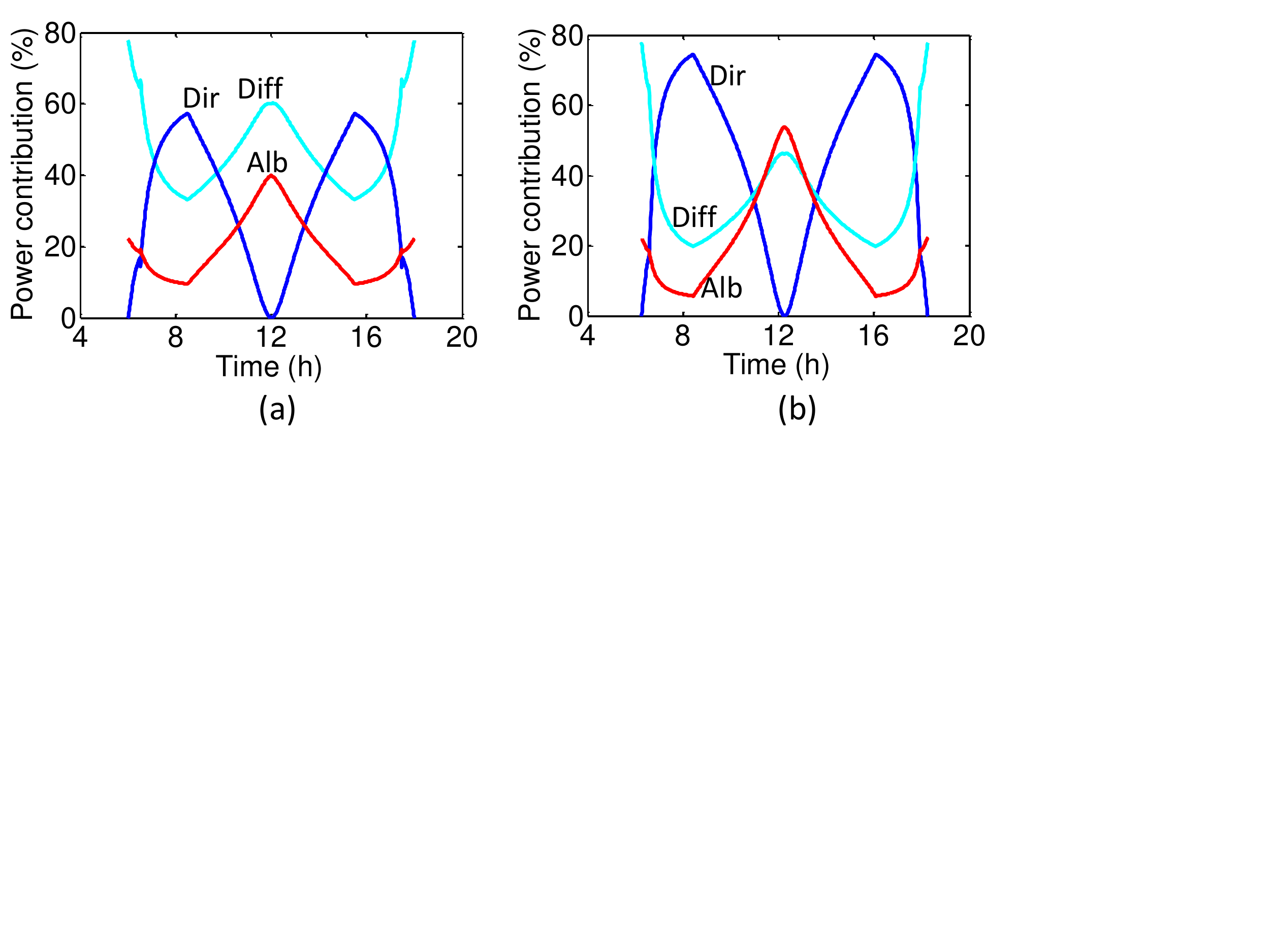}
\vspace{-125pt}
\caption{ Fractional panel-illumination contribution of direct, diffuse, and albedo light on September 22 in (a) Washington DC, and (b) Jeddah. }
\label{fig:result_dailyComp}
\end{figure}


\section{Results and Discussion}

\subsection{Hourly energy output}
\label{sec:physicsResult}

For the following calculations, we assume a typical panel height of  $h=1.2$m and the albedo reflectance of 0.5. Albedo reflectance of 0.5 or more is observed naturally for snow-covered ground, or can be achieved artificially for example by white concrete \cite{noauthor_calculating_2016}.

As discussed earlier, the hourly insolation and power generation from a farm (with period $p=2$m, i.e., $p/h=1.667$) is shown in Fig. \ref{fig:daily_power} for Washington DC (September 22). The fractional contribution of each component (direct, diffuse, and albedo) provides additional information about how the vertical bifacial panel behaves under various weather conditions. For example, in September, the insolation in Washington DC is more diffuse compared to Jeddah. Fig. \ref{fig:result_dailyComp} show the hourly fractional generation from the three components for Washington DC and Jeddah, respectively. We observe that the fractional contribution from diffuse and albedo light peaks at noon. In Washington DC, diffuse and direct components have similar contributions in early part of the day (8-10h). On the other hand, the generation from diffuse light is much lower than direct in early (8-10h) or late (14-16h) part of the day in Jeddah---this indicates that Jeddah has a more clear sky (i.e., mainly direct light). This will affect the net power production at noon. As there is no contribution from direct light at noon,  a higher fraction of diffuse light can even out the hourly output variation. 

As we will see later, the output varies as a function of the $p/h$-ratio. Therefore, the discussions above hold for any $h$ while $p/h=1.667$ is maintained.


\begin{figure}[t]
\vspace{-0pt}
\centering
\includegraphics[width=0.55\textwidth]{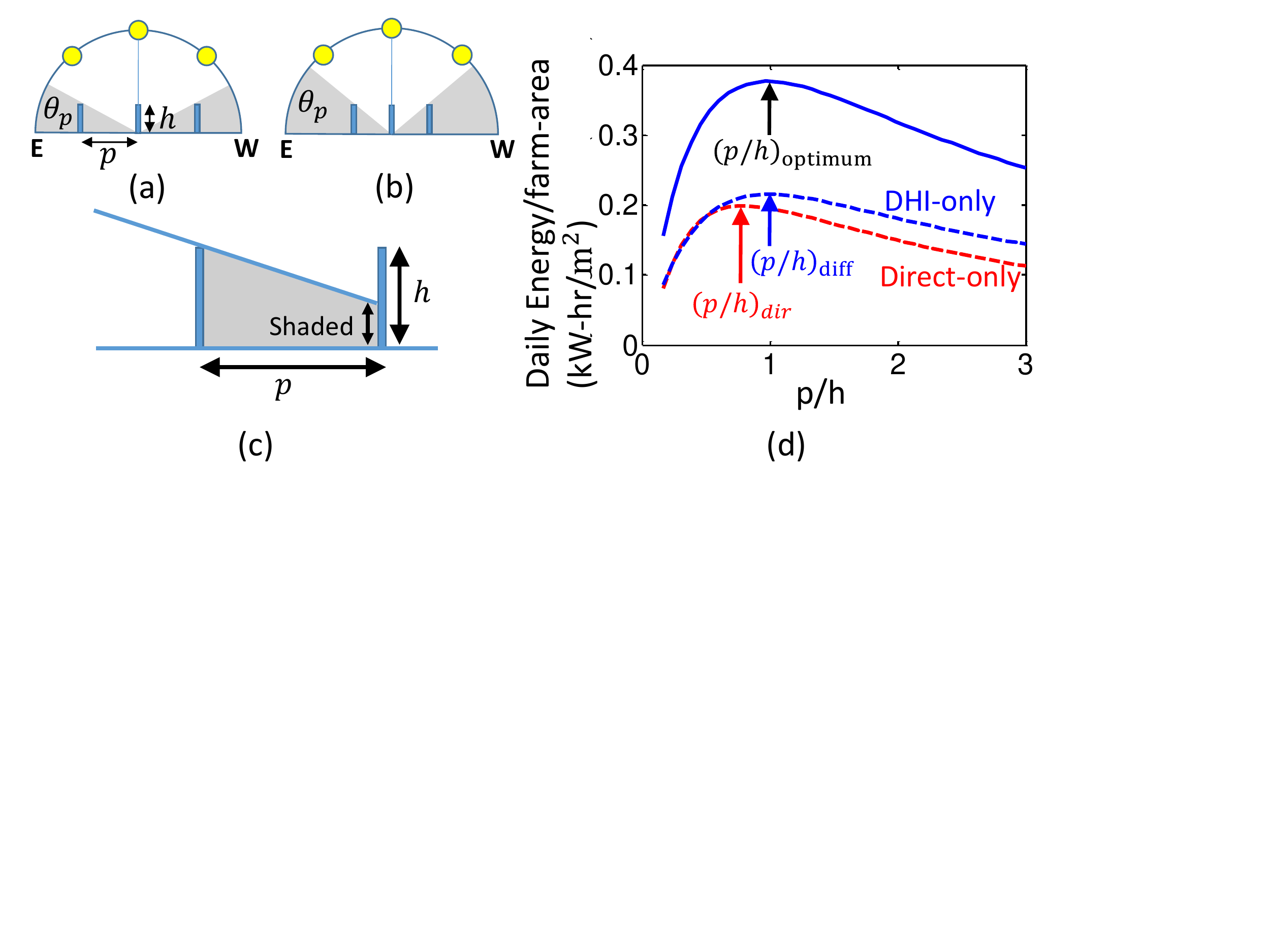}
\vspace{-120pt}
\caption{Low elevation of the sun in early morning and late evening causes mutual shading between adjacent panels. The shaded region in (a, b) indicates the time or solar positions when there is mutual shading. Clearly, larger panel-gap $p$ will have shorter shading time as in (a). (c) shows partial shading of a panel for early morning. (d) A single day energy output per farm area is shown (Sept. 22) by the blue solid line. The blue and red dashed line corresponds to the cases when only diffuse sunlight (DHI-only), and only direct sunlight (Direct-only) is considered respectively.  }
\label{fig:result_opt}
\end{figure}


\subsection{Effect of panel array period $p$}

Next let us consider the effect of the period $p$ on the farm output. Due to the array configuration, front-face of a  panel is partially illuminated (partially shadowed) in the early part of the day. For example, in Fig. \ref{fig:result_opt}(c), we see the bottom part of the panel is shaded when sun's elevation is low. In this situation, the bypass diode will turn-off the bottom string of the panel, thereby limiting the output from only the top part of the panel. Similar situation occurs for the back-face of the panel before sunset. The shadow-limited-operating conditions are shown as gray-shaded region in Fig. \ref{fig:result_opt}(a, b). 

  When the panels are packed close (i.e., small $p$), the panels on the farm have bypass-diode limited operation for a long period of each day---this greatly reduces power generation compared to light collection. Again, at large $p$, the output of each panel saturates (to ``standalone'' panel limit), and thus the farm output per unit area will decrease with increasing  $p$. Therefore, there is an optimum $p$ for which the power output \textit{per land area} is maximized, also shown by the blue solid line in Fig. \ref{fig:result_opt}(d). The optimum $p$ scales proportionally with $h$, i.e., universality of the design holds for the $p/h$ ratio. The universality may be understood by realizing that all the expressions for insolation collection contain the ratio $p/h$.
For a farm design, instead of integrating over a single day, the output is integrated over the whole year to find a functional relation similar to the one shown in  Fig. \ref{fig:result_opt}(d) for `net annual energy versus $p/h$'. This allows us find maximum annual yield and the corresponding optimum $p/h$ for that specific location. Subsequently, the analysis is repeated for various locations across the globe and a map of the location-specific optimum $p/h$ is shown in Fig. \ref{fig:result_global}(b).    The worldwide optimum $p/h$ will be discussed in the next section. It is important to highlight that energy per land area is but one metric of optimization. A levelized cost of electricity (LCOE) optimization will be a part of a future study, but we believe the key conclusion will remain the same.

\begin{figure}[t]
\vspace{-0pt}
\centering
\includegraphics[width=0.65\textwidth]{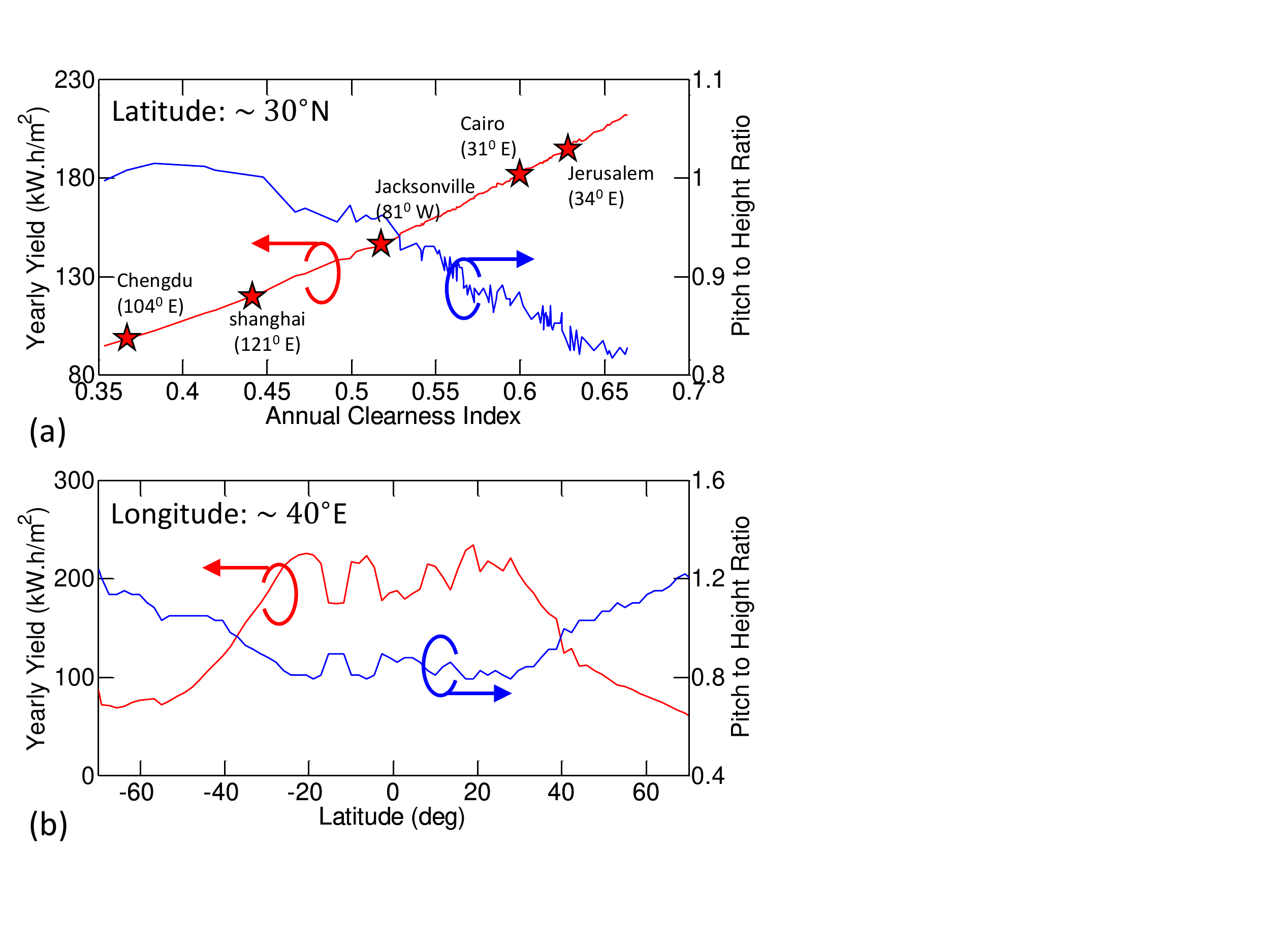}
\vspace{-30pt}
\caption{ Annual yield and optimum $p/h$ as function of (a) annual clearness index $\bar{k}_T$ (latitude $30^\circ$N), and (b) latitude (at longitude $40^\circ$E). }
\label{fig:result_kT_lat}
\end{figure}


\subsection{Effect of clearness index and latitude.}
\label{sec:kTResult}

At a given latitude on Earth, the tilt of the sun and the sun-path is the same for all longitudes. Ideally, we can thus expect the insolation to depend only on latitude. However, variation in meteorological conditions over longitudes at a set latitude causes variation in GHI, clearness index, and the fractional contribution of diffuse insolation. Such variation in weather affects the optimal design of the vertical bifacial solar farm and its yearly yield. For example, in Fig. \ref{fig:result_kT_lat}(a) we present the optimum $p/h$ (blue) and the corresponding yearly output (red line) of the farm as a function of annual clearness index $\bar{k}_T$ at latitude $30^\circ$N---the locations corresponding to the various $\bar{k}_T$ are also marked in the figure. Here, we optimize $p/h$ to maximize the annual yield, and calculate the corresponding farm output for different longitudes, but  at a fixed latitude $30^\circ$N. Then the results are sorted as a function of corresponding $\bar{k}_T$ to obtain the plots shown in Fig. \ref{fig:result_kT_lat}(a).  

At a set latitude, the sun-path is fixed (for all longitude), which in turn determines the panel shadow length and dominates the choice of optimum $p/h$. However, as shown in Fig. \ref{fig:result_kT_lat}(a), at a fixed latitude $30^\circ$N, there is a small variation in optimal $p/h$ with $\bar{k}_T$. 
In order to understand why $p/h$ decreases with $\bar{k}_T$, we need to explain the relative roles of diffuse and direct light. In Fig. \ref{fig:result_opt}(d), the blue and red dashed lines consider solar sources when only the diffuse and direct light are present, respectively. 
For the `Direct-only' case (red dashed line), the diffused light contribution is set to zero and only the direct light and its corresponding albedo contributions are used to calculate the energy output. The output is maximized at $(p/h)_\textit{dir}$, as marked by the red arrow. Similarly, for the `DHI-only' case (blue dashed line), the direct sunlight is set to zero,  and the diffused light and its albedo contributions are accounted for. The `DHI-only' contribution maximizes at $(p/h)_\textit{diff}$, as marked by the blue arrow. In general, for any location, we find that $(p/h)_\textit{dir}<(p/h)_\textit{diff}$. 
Therefore, when $\bar{k}_T$ is low, DHI or diffuse component dominates and the overall optimum $p/h$ converges to $(p/h)_\textit{diff}$. For example, note that the overall peak $(p/h)_\textit{optimum}$ (on the blue solid line in Fig. \ref{fig:result_opt}(d)) is close to $(p/h)_\textit{diff}$ position. Therefore, as shown in Fig. \ref{fig:result_kT_lat}(a), reducing $\bar{k}_T$ below 0.45 will only keep the $p/h$ at a constant value, close to $(p/h)_\textit{diff}$, dictated by the diffuse light component. In contrast, as $\bar{k}_T$ increases, the direct light starts to dictate the farm output, and the optimum $p/h$ decreases from $(p/h)_\textit{diff}$ and shifts towards $(p/h)_\textit{dir}$. Obviously, by definition, increasing $\bar{k}_T$ increasing GHI. Therefore, the optimum yearly yield increases with $\bar{k}_T$.

The locations with higher latitudes sees larger tilt in the sun-path, and lower GHI. Therefore, the yearly output is high close to the equator and decreases at higher latitudes, as shown by red line in Fig. \ref{fig:result_kT_lat}(b). From equator up to latitude $\sim 30^\circ$, the optimum $p/h$ remains close to 0.8 (blue line), and then it increases. For higher latitudes, the tilt of the sun (i.e., $\theta_Z$) is larger. The longer shadows results in higher spacing between the panels.

\subsection{Global Map of Energy Yield}
\label{sec:globalResult}

\begin{figure}[t]
\vspace{-0pt}
\centering
\includegraphics[width=0.5\textwidth]{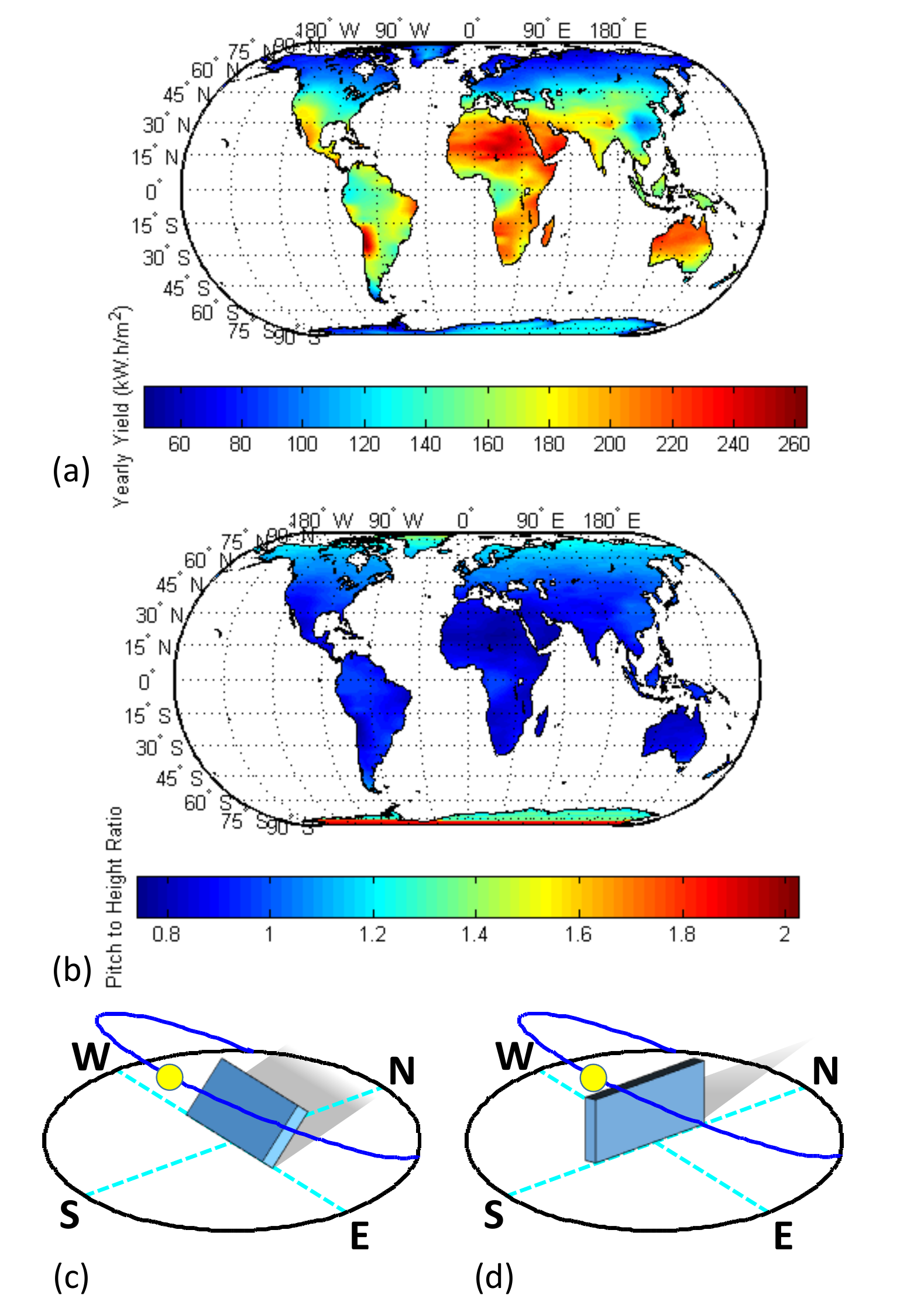}
\vspace{-10pt}
\caption{Global optimized (a) yearly yield, and (b) optimum $p/h$ for the vertical bifacial farm.  We have chosen constant 50\% albedo reflection. The sun-path is shown for latitude $70^\circ$N (July) in (c) with South facing monofacial and (d) with East-West facing vertical bifacial panel.  }
\label{fig:result_global}
\end{figure}

We are now ready to summarize the global optimization and energy yield of  vertical bifacial solar farms. We assume a constant ground reflectance of 0.5. As explained earlier, we expect decrease in GHI and output for increasing latitudes. And, there are variation in design and output along the longitudes due to meteorological variations. The global yearly yield and the corresponding optimum $p/h$ is shown in Fig. \ref{fig:result_global}. We observe higher output in Africa and Saudi Arabia compared to India and China due to clearer sky (i.e., higher $\bar{k}_T$) and higher GHI. Also, optimum $p/h \sim 0.8$ is close to equator, and begins to increase above $30^\circ$ latitude. {\it The optimum $p/h$ is  within 0.8-1 for most of the locations in the world.}


We also compare the vertical bifacial solar farm with the optimized monofacial (optimized tilt \cite{luque_handbook_2011} and spacing) solar farm. Conventionally, in a monofacial solar farm, the row spacing is selected such that the annual shading loss is less than 5\%. This results in a set-back-ratio of 2 closer to equator, and 3 for mid latitudes \cite{luque_handbook_2011}. For the monofacial farms, we take into account the direct and diffuse insolation, and neglect the albedo.  We will compare two cases: when the spacing is optimized for maximum energy yield vs. when the spacing is fixed by practical considerations.  Note that the comparison is somewhat biased because while the monofacial farm is tilt-optimized, the vertical farm -- by definition -- is not.

\begin{figure}[t]
\vspace{-0pt}
\centering
\includegraphics[width=0.85\textwidth]{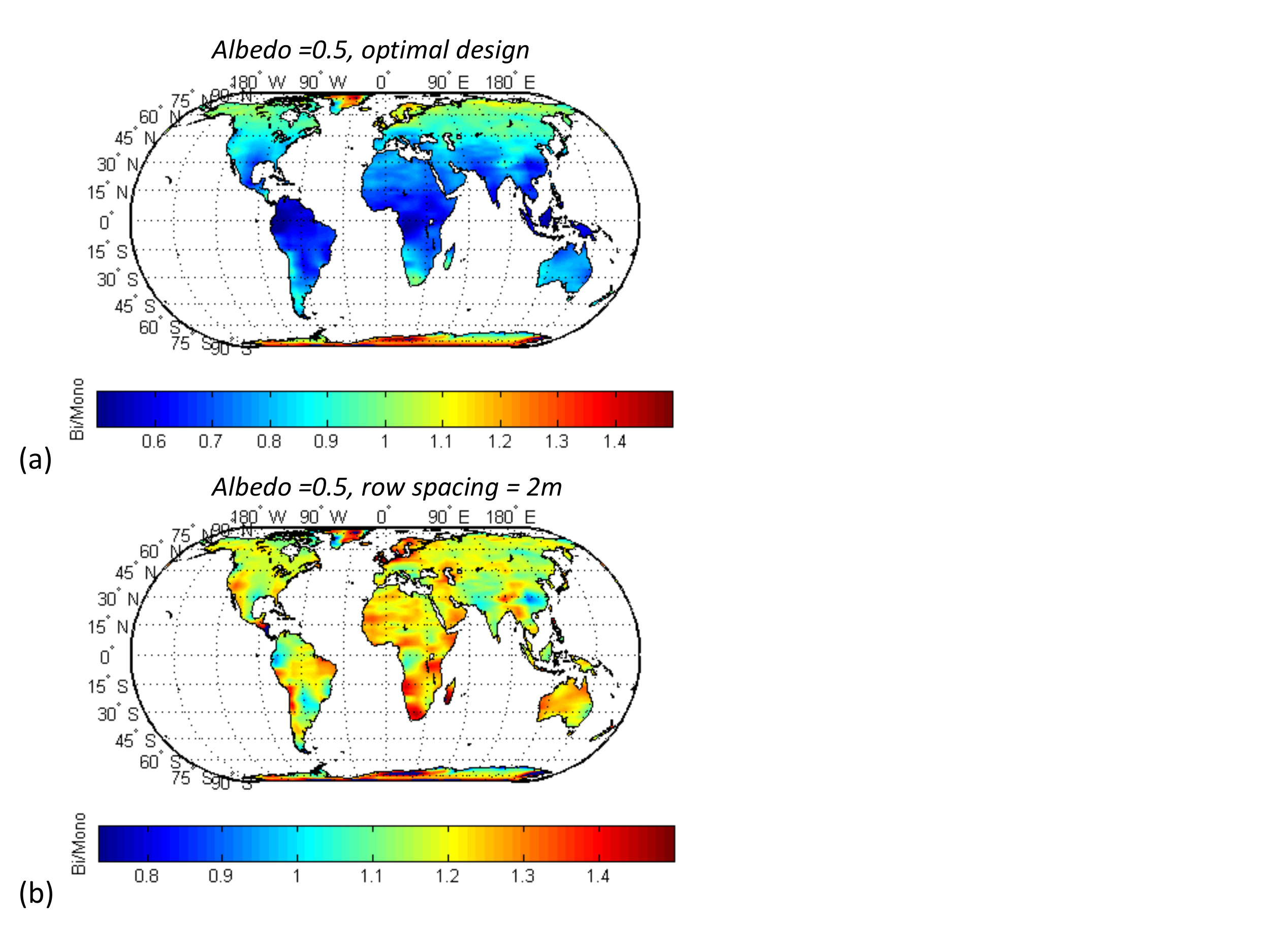}
\vspace{-35pt}
\caption{ Ratio of yearly yield from  the vertical bifacial farm to monofacial farm for (a) optimal design, and (b) fixed row spacing of 2m (1.2m wide panels). In (a), the monofacial farm has optimum  panel tilt and row spacing at each location.  }
\label{fig:result_global_ratio}
\end{figure}

\noindent{\bf 3.4.1 Spacing optimized solar farms.} 
For the first  comparison, we  determine the optimum row spacing to  maximize the area-normalized annual energy yield of monofacial and bifacial farms for a given location in the world.  The ratio of the `maximized' annual yield of the vertical bifacial  farm to the monofacial farm is plotted in Fig. \ref{fig:result_global_ratio}(a). Close to equator, the monofacial panels are optimally tilted parallel to the ground, and the optimal row spacing is close to zero. Close to equator, therefore,  monofacial panels collect the GHI fully, yielding the maximum output for {\it any} farm configuration. In these locations, in absence of any soiling considerations, this energy output is twice as large compared to a vertical bifacial farm. The advantage of monofacial farms decreases at higher latitude. 
At latitudes $>60^\circ$, the sun-path is highly tilted. For example, consider the sun-path shown in Fig. \ref{fig:result_global}(c,d) for latitude $70^\circ$N in July. The sun is at North-East (or North-West) in early (or late) part of the day. At these times, the South-facing monofacial panel does not receive any direct sunlight (Fig. \ref{fig:result_global}(c)), unlike East-West facing bifacial panel (Fig. \ref{fig:result_global}(d)). Moreover, closer to noon, when the insolation is more significant, we would observe a long shadow towards the North. This result in prominent shading on adjacent South-facing monofacial panels, see Fig. \ref{fig:result_global}(c). The shadows towards the East or West are relatively shorter; therefore, the East-West facing vertical bifacial panels incur lower shading loss. The bifacial panels allows the vertical farm to collect more energy both from the sky and the ground compared to the optimally (and highly) tilted monofacial panel array. In these locations at high latitudes, the bifacial farm produces significantly more energy than monofacial farms.  

\noindent{\bf 3.4.2 Spacing with practical considerations.} Recall that the period $p$ of an array is defined by the sum of the row spacing and the horizontal distance covered by a tilted panel. Unfortunately, close to equator (within $30^\circ$ latitudes), yield-optimized monofacial farms have a row-spacing less than $0.25$m, which is impractically small  for installation and maintenance purposes.  For example, the row-spacing in the farm is required to be 2m or higher \cite{lindsay_key_2015}. Therefore, the `yield-optimized' comparison of the farms close to equator (as discussed in the preceding section) may not be practical. Therefore, next we compare the energy yield of the  farms with fixed 2m row-spacing.

The ratio of the annual yield of the vertical bifacial  farm to the monofacial farm, assuming 2m row-spacing for all, is shown in Fig. \ref{fig:result_global_ratio}(b). The conclusion is obvious: For almost all regions of the world, ground-mounted vertical bifacial farms outperform tilt-optimized monofacial farms by 10-20\%. Indeed, some regions in Africa and South America may offer  50\% more energy output. However, there are a few isolated locations in the world (e.g., parts of China, Columbia, Equador, etc.--marked in deep blue in Fig. \ref{fig:result_global_ratio}(b)) where bifacial cell under-performs a monofacial cell by 10-20\%. These regions are characterized by low clearness index, so that the shading of the diffuse light at the bottom of the panel and the current-constraint associated with the lower bypass diode strongly penalizes the power output of a bifacial farm (see the corresponding regions in Fig. \ref{fig:result_global}(a)). In these regions, bifacial farms may only be viable if the panels are optimized for tilt angle and energy-penalty due to soiling are accounted for. Indeed, vertical farms seem even more attractive as cleaning costs (e.g., water, labor), etc. are expected to be lower and overall reduction in temperature will improve farm operating lifetime. Therefore, a LCOE-based optimization is essential to accurately quantify the possible gain in utilizing the vertical bifacial farm. 


\section{Summary and Conclusions }
\label{sec:conclusion}

In this paper, we have utilized worldwide meteorological data and a detailed physical panel-array model to estimate annual yield of vertical bifacial solar farms. To summarize:
\begin{enumerate}
\item We have combined the daily average meteorological NASA data \cite{power_surface_2017}  with a \textit{clear-sky model} from Sandia \cite{sandia_national_labs_pv_2016,haurwitz_insolation_1945,haurwitz_insolation_1946} to obtain hourly insolation information. This combined model greatly reduces loading of large database and speeds up computation while maintaining average meteorological insolation information.

\item Our physics model for panel array combines the effect of direct, diffuse, and albedo illumination onto the panels. The mutual shading and collection from direct and diffuse insolation has been modeled based on previous literature. We have discussed partial shading and illumination onto the ground between panels due to direct and diffuse sunlight. The non-trivial collection of this albedo light has been explained in details. The spatially non-uniform illumination along the panel height can affect the final panel output---our general formulation of the model allows us to calculate such details. 

\item Due to non-uniform illumination, and string of series-connected cell configured on a panel, the energy output is not proportional to the insolation collection. We assume that the panel is divided into 3 sub-strings each connected with its own bypass-diodes. The final output of the panels are calculated for this specific configuration.   

\item Mutual shading between adjacent panels restricts panels being closely packed in the farm. We explain how this results in an optimum period between the panels. At high latitudes, the sun-path is more tilted, resulting in larger optimum panel-period. And, at the same latitude, locations with more diffuse insolation (i.e., lower $\bar{k}_T$) tend to have a larger panel-period.

\item We present a global perspective on the annual yield of vertical bifacial solar farms. For a practical row-spacing of 2m, the energy yield of bifacial solar farms continues to outperform the monofacial solar cells in most of the regions of the world, although the energy gain is somewhat smaller compared to stand-alone panels  \cite{guo_vertically_2013}.

\end{enumerate}

Finally, we wish to highlight three factors that will impact the LCOE of a solar farm, but were deemed beyond the scope of the paper. First, the increased energy yield of a bifacial farm requires closely spaced panels. Since the bifacial panels are somewhat more expensive, the LCOE must be calculated carefully to reflect this additional cost. Second, a recent study shows that vertical panels have low dust accumulation while having energy yield similar to conventional tilted panels \cite{hajjar_bifacial_2015}, because of the {\it soiling penalty} associated with the monofacial cells. Moreover, cleaning the panels is expensive. Therefore, the energy gain of vertical farms, in practice, may be higher than those summarized in Fig. \ref{fig:result_global_ratio}. Finally, a farm designed with optimally tilted and {\it elevated} panel array produces much more energy than a ground-mounted vertical bifacial farm \cite{appelbaum_bifacial_2016}. Overall, these energy gains must be balanced carefully with the increased installation cost to ensure the worldwide economic viability of the bifacial solar farms.





\section*{Acknowledgement}

We gratefully acknowledge Dr. C. Deline from NREL who read the initial draft of the paper and highlighted the importance of bypass diodes regarding the energy yield of bifacial farms. This work was made possible through financial support from the National Science Foundation through the NCN-NEEDS program, contract 1227020-EEC and by the Semiconductor Research Corporation, the US-India Partnership to Advance Clean Energy-Research (PACE-R) for the Solar Energy Research Institute for India and the United States (SERIIUS), U.S. Department of Energy under Contract No. DE-AC36-08GO28308 with the National Renewable Energy Laboratory, the Department of Energy under DOE Cooperative Agreement No. DE-EE0004946 (PVMI Bay Area PV Consortium), and the National Science Foundation under Award EEC1454315-CAREER: Thermophotonics for Efficient Harvesting of Waste Heat as Electricity.

\bibliography{Zotero}

\end{document}


\begin{frontmatter}

\title{Supplementary Information:\\ Vertical Bifacial Solar Farms: Physics, Design, and Global Optimization}

\author[purdue]{M. Ryyan Khan  \fnref{eq_contr} }
\author[kaust]{Amir Hanna  \fnref{eq_contr} }
\author[purdue]{Xingshu Sun  \fnref{eq_contr} }
\author[purdue]{Muhammad A. Alam\corref{email_Alam}\fnref{eq_contr}}\ead{alam@purdue.edu}

\address[kaust]{Electrical Engineering Department, King Abdullah University of Science and Technology, Saudi Arabia}

\address[purdue]{School of Electrical and Computer Engineering, Purdue University, West Lafayette, IN 47906, United States}

 \fntext[eq_contr]{The authors contributed equally.} 
\cortext[email_Alam]{Corresponding author}

\end{frontmatter}

\section{Panel efficiency under uniform illumination}

The efficiency $\eta_{\textit{diff}}$ of the panel under isotropic illumination can be estimated from the weighted average of $\eta(\theta)$ over all the solid angles. However, we calculate $\eta$ and $\eta_{\textit{diff}}$ for a realistic panel using the simulator `Tracey' \cite{keith_r._mcintosh_tracey_2013,mcintosh_optical_2009,mcintosh_optical_2010}.  The simulator does ray tracing for a panel structure considering interface reflections, scattering from metal fingers, and losses in glass and encapsulation. We assume modules that have square cells (15.6 cm). There is 3 busbars per cell, where the busbar width is 0.15 cm and the bus finger pitch is 0.2 cm. The bus finger thickness is 150 $\mu$m. The thicknesses of the encapsulation (EVA) and glass layers are 450 $\mu$m and 0.2 cm, respectively. For simplicity, the edge losses are neglected in the efficiency calculation.

We simulate two specific cases: (i) direct light incident normal to the panel, (ii) isotropic (diffused) light incident on the panel. We assume illumination on a single face of the panel. The ray tracer finds light absorption and combines the result with a typical internal quantum efficiency (IQE) of a Si-panel to find short circuit current. We assume  open circuit voltage of 0.68V, and 78.5\% fill factor (FF)---this yielded in $\eta=18.9\%$ and $\eta_{\textit{diff}}=15.67\%$ under normal and diffused illumination on a single face of the panel. 

\section{Panel operation under partial shading}

\subsection{Panel configuration}

\begin{figure}[t]
\vspace{-0pt}
\centering
\includegraphics[width=0.55\textwidth]{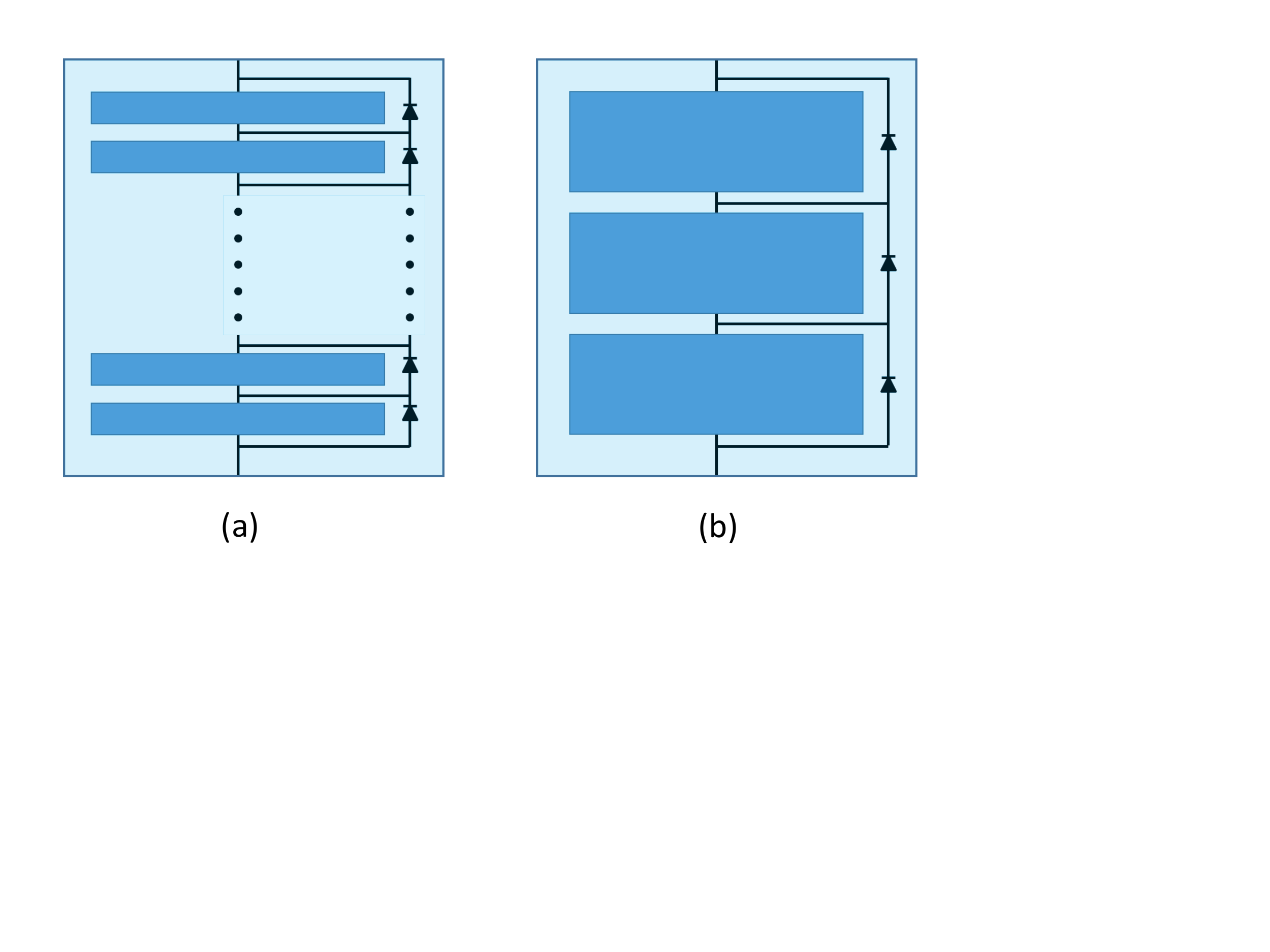}
\vspace{-110pt}
\caption{ (a) A schematic configuration of the panel. Each sub-string (blue block)  consists of several series connected cells (not shown separately), and has a bypass-diode in parallel. There are $N_{bypass}$ sub-strings in series. (b) A specific case when $N_{bypass}=3$. }
\label{fig:panel_config}
\end{figure}

The panel is composed of series connected cells. The full configuration is assumed to be divided into $N_{bypass}$ rows of sub-strings, see Fig. \ref{fig:panel_config}(a). Each sub-string is protected by a bypass-diode. For the calculation reported in this paper, we assume $N_{bypass}=3$ as shown in Fig.~\ref{fig:panel_config}(b). With a panel height of $h$, each sub-string section has a height of $h_N=h/N_{bypass}$.


\subsection{Non-uniform illumination}

Partial shading creates non-uniform illumination across the panel, with the bottom receiving less energy compared to the top. The statement is obvious for the direct light As discussed in the main manuscript, shadow height on the bottom of the panel is $h_s$---this height will change throughout the day as the position of the sun changes, indicating that the lower part of the panel receives less energy than the top. 

The considerations are slightly more complicated for diffuse and albedo light. Similar to the direct light, the diffuse insolation provides higher intensity towards the top-edge of the panel compared to the bottom-edge due to partial shading between panels. On the other hand, the scattered albedo provides higher intensity close to ground (bottom-edge) compared to the top-edge of the panel. Despite this counter-acting intensity profiles of diffuse and albedo lights, our calculations show that for ground reflectance of 0.3-0.6 and for all practical $p/h$ and clearness indices, the combined  illumination from diffuse and albedo light yields in a profile with the lowest intensity towards the bottom of the panel.

\subsection{Panel output calculation steps}

We utilize the analytical model developed in Ref. \cite{deline_simplified_2013} to find how the  panel configured with 3-bypass diodes respond to non-uniform illumination. The model assumes diffuse light (diffuse insolation and albedo combined) $\widehat{I}_{PV(d,A)}^{\textit{(bifacial)}}$ uniformly illuminating the panel. This is also the lower limit to the panel output when the panel is mostly shadowed. However, as we have explained in the preceding section, the bottom section of the panel has the lowest contribution from the combined diffuse insolation and albedo. We can thus define this lower output limit from the diffused light collection at the bottom section of the panel:
\begin{align}
\widehat{I}_{PV(d,A)}^{\textit{(bifacial)}}
&= \frac{1}{h_N} \int_0^{h_N} {  \left(  \widehat{I}_{PV(diff)}^{\textit{(bifacial)}}(z) + \widehat{I}_{PV(Alb)}^{\textit{(bifacial)}}(z)  \right) dz}
\end{align}

Next, the output from direct light assuming unshaded panels (mutual shading neglected) is as follows:
\begin{equation}
\widehat{I}_\textit{PV(dir),UnSh}^{\textit{(bifacial)}} = \frac{1}{h} \int_0^h {\widehat{I}_{PV(dir)}^{\textit{(bifacial)}}(z)\, dz  }
\end{equation}
The `diffused fraction' is therefore,
\begin{equation}
f_D = \frac{\widehat{I}_{PV(d,A)}^{\textit{(bifacial)}}}{ \widehat{I}_{PV(d,A)}^{\textit{(bifacial)}}+\widehat{I}_\textit{PV(dir),UnSh}^{\textit{(bifacial)}}}
\end{equation}
Finally, the overall power reduction $f_{out}$ is found as a function of $(N_{bypass}, h_s/h, f_D, FF)$ (see Equ. (15) in Ref. \cite{deline_simplified_2013}). The output  is then:
\begin{align}
\widehat{I}_{PV}^{\textit{(bifacial)}} 
	&= f_{out} \times \left(  \widehat{I}_{PV(d,A)}^{\textit{(bifacial)}}+\widehat{I}_\textit{PV(dir),UnSh}^{\textit{(bifacial)}}  \right), \\
    I_{PV}^{\textit{(bifacial)}}  &= \frac{h}{p} \widehat{I}_{PV}^{\textit{(bifacial)}} .
\end{align}
Here, $\widehat{I}_{PV}^{\textit{(bifacial)}} $ and ${I}_{PV}^{\textit{(bifacial)}} $ are total output \textit{per panel area} and \textit{per farm area}, respectively.

\section{Global output with array period}

\begin{figure}[h]
\vspace{-0pt}
\centering
\includegraphics[width=0.65\textwidth]{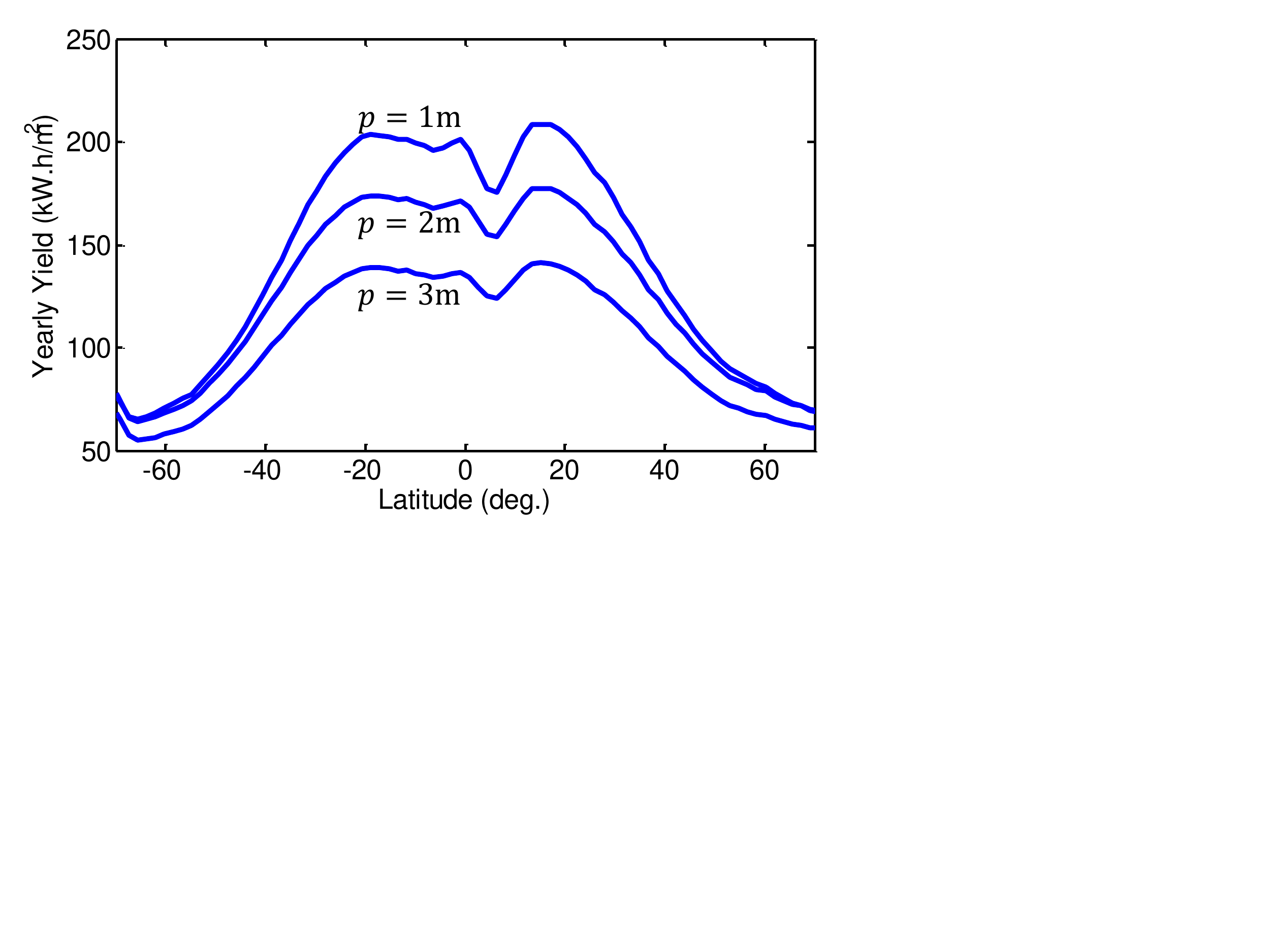}
\vspace{-125pt}
\caption{ Yearly yield (averaged over longitudes) of the vertical bifacial farm as function of latitude for set periods of $p=1,2,3$m. }
\label{fig:yearly_output_p}
\end{figure}

As discussed in the main manuscript, instead of optimizing for the period $p$ of the vertical panel array, we may need to fix $p$ for maintenance purposes. For example, Fig. \ref{fig:yearly_output_p} shows yearly yield of the farm for set $p=1,2,3$m. The optimum $p/h$ is within 0.8-1.1m (we have $h=1.2$m); therefore for larger periods (e.g., $p=2,3$m), the output will be lower, as seen in Fig. \ref{fig:yearly_output_p}.


\section{Illustrative discussion on specific locations}

In this section, we will discuss insolation and farm yield for five specific locations listed in Table \ref{tab:location_list}. The regions span across continents, have very different elevation and cloudiness (characterized by clearness index). Our goal is to illustrate the validity of the model by focusing on energy output in specific regions of the world.  The GHI and location specific elevation are found from NASA Surface meteorology and Solar Energy database \cite{power_surface_2017}. The average yearly clearness index (or annual clearness index), $\bar{k}_{T}$ is found from GHI and extraterrestrial insolation of the respective locations.

\begin{table}[h]
\centering
\caption{List of location for the illustrative discussion. The elevation values of the corresponding locations are found from the NASA-database. The yearly average clearness index is found based on the insolation model described in the main text.}
\label{tab:location_list}
\footnotesize \vspace{2pt}
\begin{tabular}{|l|l|l|p{0.06\textwidth}|}
\hline
 & (Lat., Long.) & Elevation & Avg. yearly $\bar{k}_{T}$ \\
\hline
Washington DC & 38.9N,	77.04W & 85m & 0.4853 \\
\hline
Leadville, Colorado & 39.25N, 106.29W & 2653m & 0.5751 \\
\hline
Broken Hill, Australia & 32S, 141.5E	& 253m & 0.6147 \\
\hline
Kamuthi, India & 9.35N,	78.39E & 64m & 0.5341 \\
\hline
Schipkau, Germany & 51.57N,	13.98E & 71m & 0.4209 \\
\hline
\end{tabular}
\end{table}


\begin{figure}[hb]
\vspace{-0pt}
\centering
\includegraphics[width=0.6\textwidth]{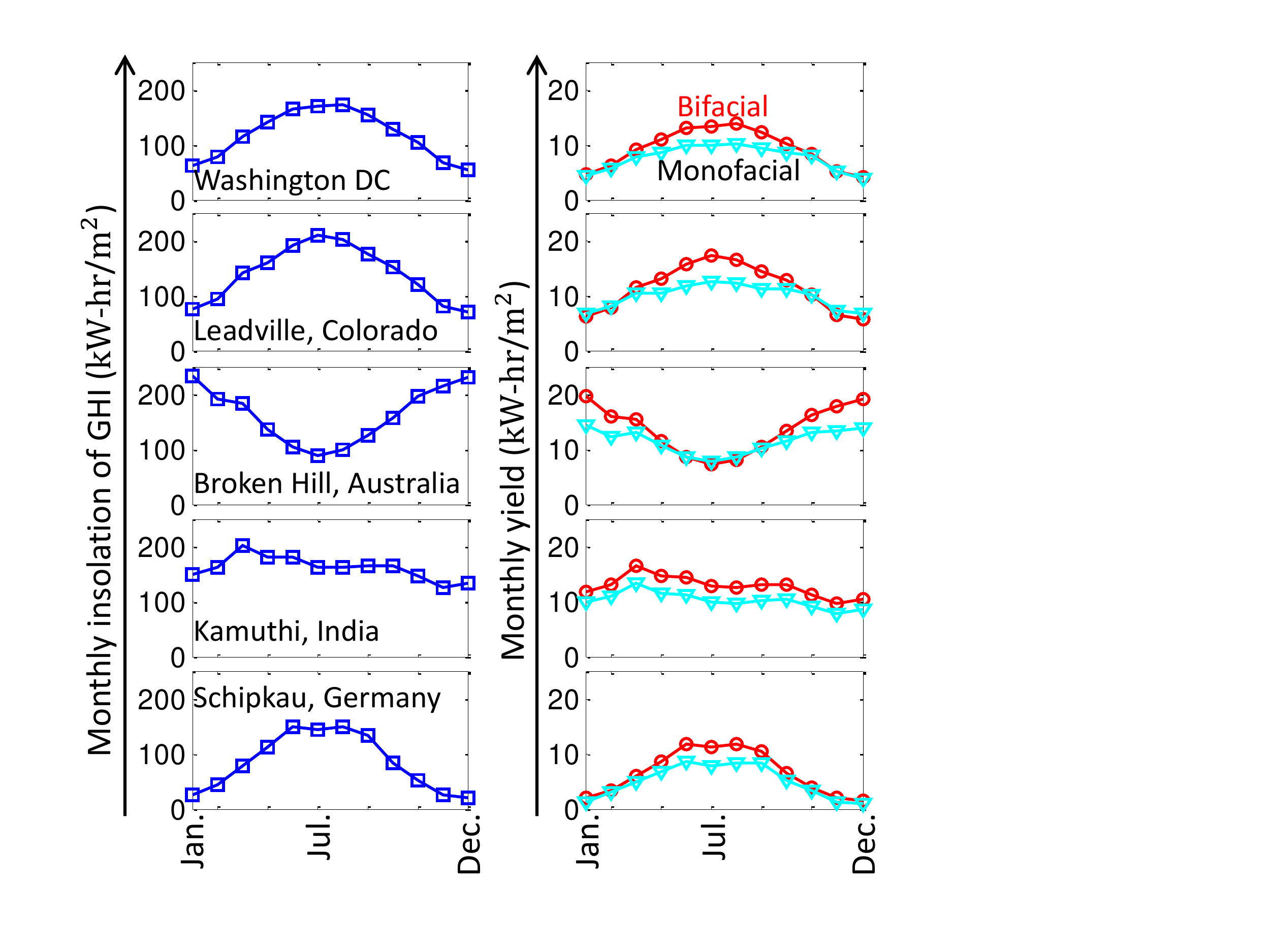}
\vspace{-30pt}
\caption{ Monthly GHI extracted from NASA-database are shown by blue squares. Assuming 2m row-spacing, calculated monthly output from vertical bifacial (red circles) and monofacial (cyan triangles) farms are also shown.}
\label{fig:monthly_output}
\end{figure}


The monthly insolation of GHI extracted from the NASA-database for each of the locations are shown by blue symbols in Fig. \ref{fig:monthly_output}. As explained in the main text, the GHI is then split into DHI and DNI based on elevation of the respective locations. Australia being in the Southern hemisphere, has lowest GHI in June (winter), as expected. These insolation information are then used to calculate farm output. Monthly yield of bifacial (red circles) and monofacial (cyan triangles) farms assuming 2m row-spacing are shown in Fig. \ref{fig:monthly_output} for each of the locations. The monofacial panels are, of course, optimally tilted as a function of latitude (tilt $=0.69\times|$Lat$|+3.7^\circ$) \cite{luque_handbook_2011}.  For this row-spacing, vertical bifacial solar farm yields higher compared to monofacial farms.


\begin{figure}[t]
\vspace{-0pt}
\centering
\includegraphics[width=0.65\textwidth]{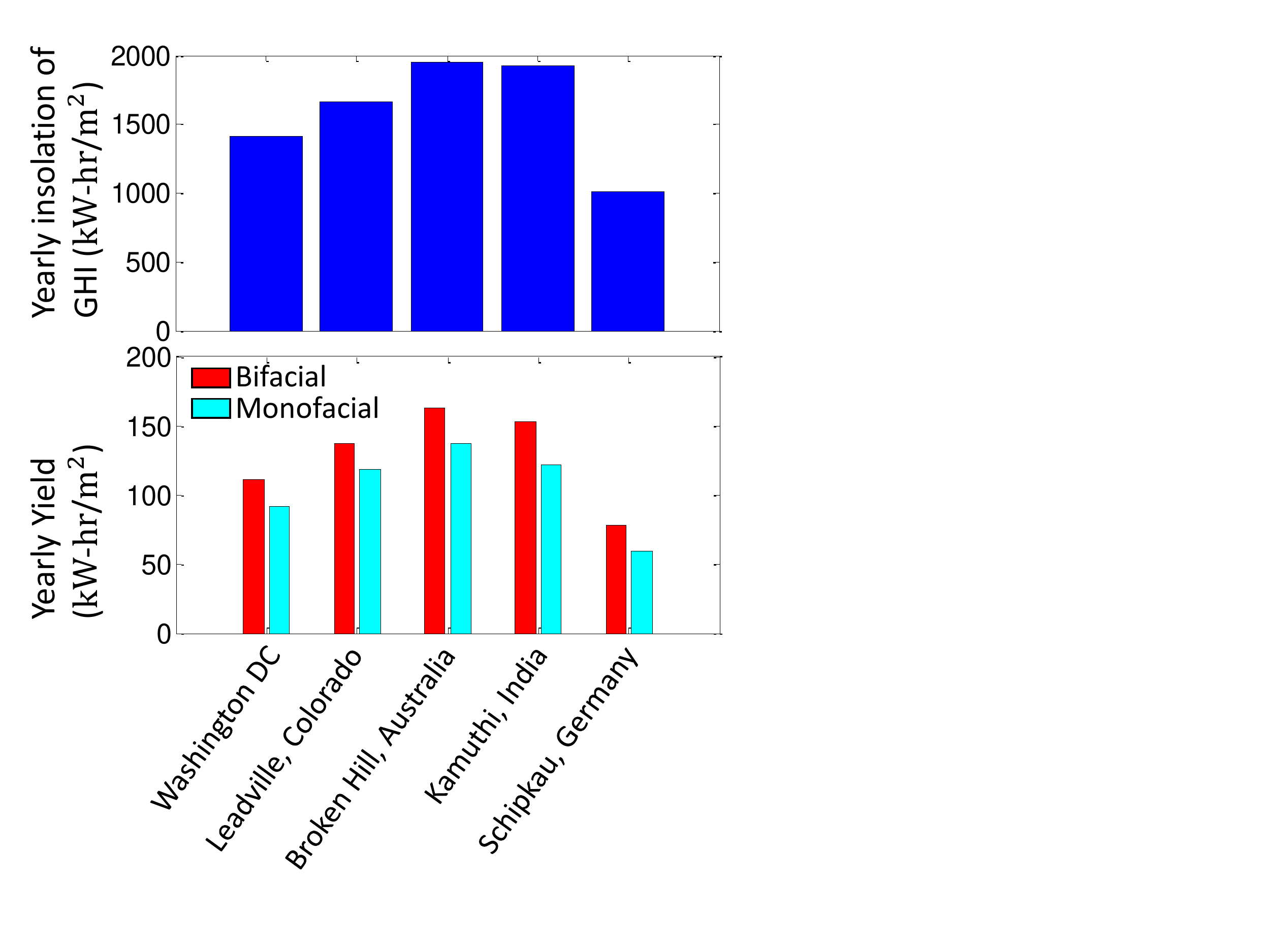}
\vspace{-40pt}
\caption{ Integrated yearly GHI, and the farm output (2m row-spacing) are shown for the locations listed in Table \ref{tab:location_list}. }
\label{fig:location_compare}
\end{figure}


The integrated yearly GHI for the locations under current study are shown in Fig. \ref{fig:location_compare}. Here, Broken Hill, Australia show the highest yearly GHI and highest $\bar{k}_T$. 
Both these values are the lowest for Schipkau, Germany. The corresponding yearly yields of the vertical and monofacial farms are also shown in Fig. \ref{fig:location_compare}. For the 2m row-spacing farm design, the vertical bifacial farms yield more compared to monofacial farms, as discussed in the main text.


\begin{figure}[b]
\vspace{-0pt}
\centering
\includegraphics[width=0.65\textwidth]{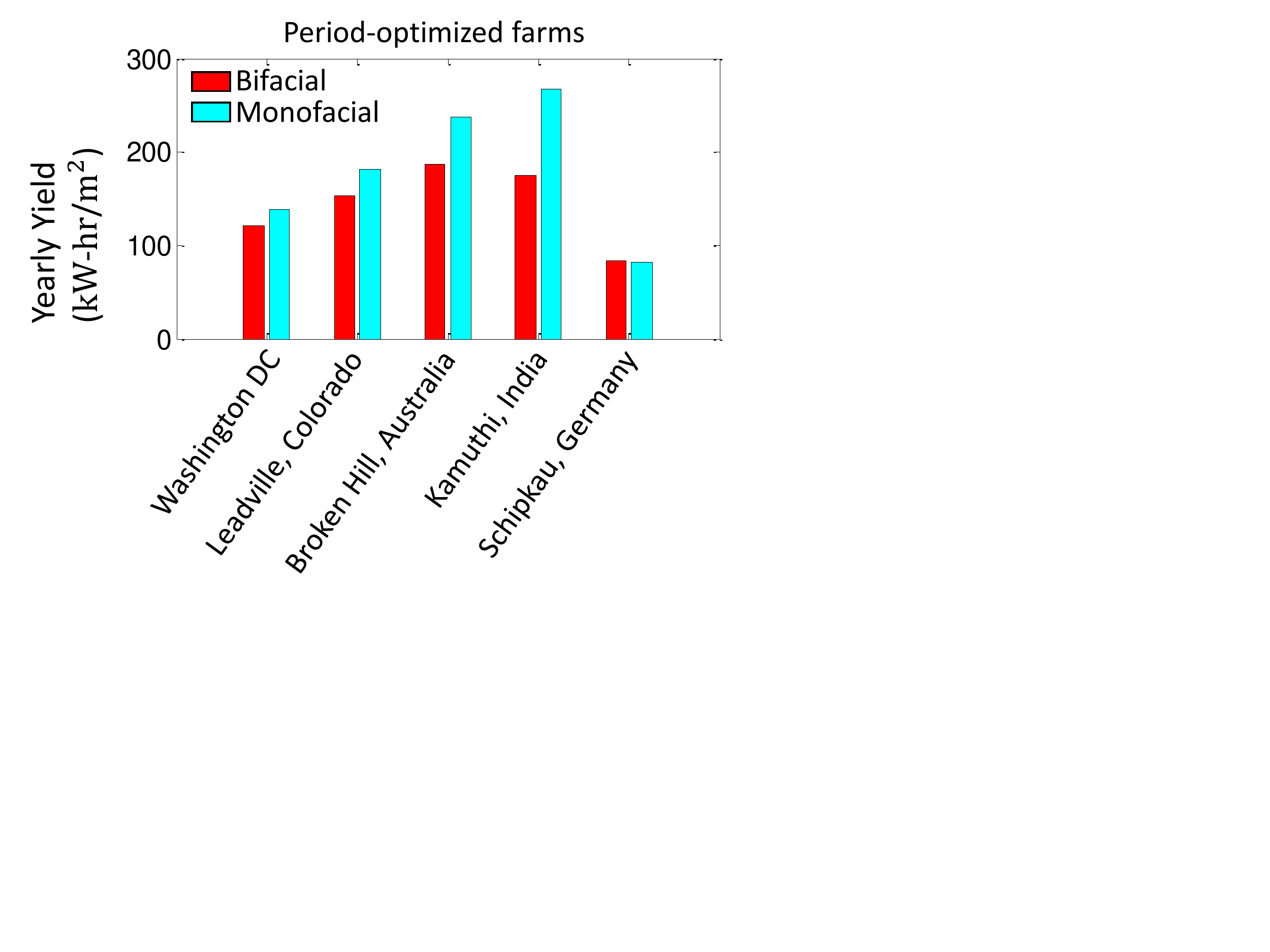}
\vspace{-120pt}
\caption{ Integrated yearly farm output (period optimized for maximum yield) are shown for the locations listed in Table \ref{tab:location_list}. }
\label{fig:location_compare_opt}
\end{figure}


Finally, the yearly yield for period-optimized farms (by maximizing output)  are shown in Fig. \ref{fig:location_compare_opt}. The bifacial farm optimizes at $p/h\sim 1$. The monofacial farm optimizes at $p/h\sim 1.1$ to 1.3 for the selected locations. This translates into very small row spacing ($<0.5$m) for $h=1.2$m, as explained in the main manuscript. However, these results in Fig. \ref{fig:location_compare_opt} indicates the maximum possible yield from vertical bifacial and monofacial farms for these locations. 


\bibliography{Zotero}